\documentclass[12pt,notitlepage]{article}
\usepackage{amsmath}         % \ 
\usepackage{amssymb}         %  | AMS Packages for math
\usepackage{amsfonts}        % /
\usepackage{graphicx}        % Graphics  

\usepackage{color}         % \color{...}, colored text
\usepackage{slashed}       % \slashed{k}
\usepackage{framed}        % boxed remarks
\usepackage{subcaption}    % subfigures; subfig depreciated
\usepackage[mathscr]{euscript} % nicer Weinberg letters
\usepackage{cite}          % grouping citations (incompatible with collref)
\usepackage{booktabs}      % tables
\usepackage{youngtab}	    % Young Tableaux
\usepackage{arydshln} 	    % dashed lines in arrays and tables
\usepackage{tocloft}		   % Table of contents
\usepackage{setspace}	   % for spacing, e.g. of TOC
\usepackage{listings}       % \begin{lstlisting}, for code
\usepackage[margin=2cm]{geometry}   % reasonable margins
\graphicspath{{figures/}}	        % set directory for figures
\numberwithin{equation}{section}    % set equation numbering
\renewcommand{\tilde}{\widetilde}   % tilde over characters
\newcommand{\email}[1]{\href{mailto:#1}{#1}}

\newenvironment{institutions}[1][2em]{\begin{list}{}{\setlength\leftmargin{#1}\setlength\rightmargin{#1}}\item[]}{\end{list}}

\let\oldenumerate\enumerate
\renewcommand{\enumerate}{
  \oldenumerate
  \setlength{\itemsep}{1pt}
  \setlength{\parskip}{0pt}
  \setlength{\parsep}{0pt}
}

\let\olditemize\itemize
\renewcommand{\itemize}{
  \olditemize
  \setlength{\itemsep}{1pt}
  \setlength{\parskip}{0pt}
  \setlength{\parsep}{0pt}
}

\usepackage{bbm}
\usepackage{fancyhdr}		% to put preprint number

\usepackage[
	colorlinks=true,
	citecolor=black,
	linkcolor=black,
	urlcolor=blue,
	hypertexnames=false]{hyperref}

\newcommand{\TeV}{\ensuremath{\text{\small TeV}}}

\newcommand{\susy}{\ensuremath{\textsc{susy}}}

\newcommand{\abs}[1]{\ensuremath{\left|#1\right|}}
\newcommand{\ZZ}{\ensuremath{\mathbbm{Z} }}

\newcommand{\Pf}{\ensuremath{\text{Pf}\,}}

\newcommand{\barB}{\ensuremath{\overline{B}}}
\newcommand{\barK}{\ensuremath{\overline{K}}}
\newcommand{\barQ}{\ensuremath{\overline{Q}}}
\newcommand{\barZ}{\ensuremath{\overline{Z}}}

\newcommand{\ff}{\ensuremath{\mathcal F}}

\newcommand{\nn}{\ensuremath{\mathcal N}}

\newcommand{\uu}{\ensuremath{\mathcal U}}

\newcommand{\ww}{\ensuremath{\mathcal W}}

\newcommand{\tJ}{\ensuremath{\tilde{J}}}
\newcommand{\tK}{\ensuremath{\tilde{K}}}
\newcommand{\tM}{\ensuremath{\tilde{M}}}

\newcommand{\tZ}{\ensuremath{\tilde{Z}}}
\renewcommand{\P}{\barQ}
\newcommand{\arr}{\ensuremath{r}}
\newcommand{\bG}{\ensuremath{\tilde{G}}}
\newcommand{\bZ}{\ensuremath{\overline{Z}}}
\newcommand{\bL}{\ensuremath{\overline{\Lambda}}}

\newcommand{\Zeven}{\ensuremath{Z_\text{even}}}
\newcommand{\Zodd}{\ensuremath{Z_\text{odd}}}
\newcommand{\bZeven}{\ensuremath{\overline{Z}_\text{even}}}
\newcommand{\bZodd}{\ensuremath{\overline{Z}_\text{odd}}}

\newcommand{\Tr}{\ensuremath{\text{Tr\,}}}

\newcommand{\yI}{\ensuremath{{\bf 1}}}
\newcommand{\yF}{\ensuremath{\tiny\yng(1)}}
\newcommand{\yFb}{\ensuremath{\tiny \overline{\yng(1)}}}
\newcommand{\yA}{\ensuremath{\tiny\Yvcentermath1 \yng(1,1)}}

\newcommand{\yAd}{\ensuremath{{\bf Adj}}}
\newcommand{\ysub}[1]{\scalebox{0.8}[0.8]{#1}}

\newcommand{\Lb}[1]{\ensuremath{ \Lambda_{#1}^b}}
\newcommand{\tLb}[1]{\ensuremath{ \tilde{\Lambda}_{#1}^b}}

\newcommand{\bLb}[1]{\ensuremath{ \overline{\Lambda}_{#1}^b}}
\newcommand{\mlr}{\ensuremath{M_{\ell\arr}}}

\newcommand{\ev}[1]{\ensuremath{ \langle #1 \rangle} }
\renewcommand{\eqref}[1]{Eq.~(\ref{#1})}
\newcommand{\Eqref}[1]{Equation~(\ref{#1})}

\fancypagestyle{firststyle}{
\rhead{\footnotesize \texttt{UCI-TR-2017-01}}}

\begin{document}

\thispagestyle{empty}
\thispagestyle{firststyle}

\begin{center}

    %{\LARGE \bf Product gauge groups and s-confinement}
    {\Large \bf Product Group Confinement in SUSY Gauge Theories}

    \vskip .7cm

    { \bf Benjamin~Lillard} 
    \\ \vspace{-.2em}
    { %\tt
    \footnotesize
    \email{blillard@uci.edu}
    }
	
    \vspace{-.2cm}

    \begin{institutions}[3.5cm]
    \footnotesize
    {\it 
	    Department of Physics \& Astronomy, 
	    University of California, Irvine.
	    }   
    \end{institutions}

\end{center}

\begin{abstract}
\noindent We propose a new set of s-confining theories with product gauge groups and no tree-level superpotential, based on a model with one antisymmetric matter field and four flavors of quarks. 
For each product group we find a set of gauge-invariant operators which satisfy the 't~Hooft anomaly matching conditions, and we identify the dynamically generated superpotential which reproduces the classical constraints between operators. Several of these product gauge theories confine without breaking chiral symmetry, even in cases where the classical moduli space is quantum-modified.
These results may be useful for composite model building, particularly in cases where small operators of the form $(Q \barQ)$ are absent, or for theories with multiple natural energy scales, and may provide new ways to break supersymmetry dynamically.
\end{abstract}

%\tableofcontents

\section{Introduction}

Experimental evidence so far suggests that the Standard Model gauge group $G_\text{SM} = SU(3)_c \times SU(2)_L \times U(1)_Y$ well describes the universe.
Attempts to expand the gauge sector beyond $G_\text{SM}$ must therefore explain why the additional interactions have not yet presented any evidence for their existence. 

There are several well-motivated ways to achieve this. The new gauge bosons and matter fields might form a ``dark sector" and interact weakly (or not at all) with the particles described by the Standard Model. It is also possible for an extended gauge symmetry to be spontaneously broken to $G_\text{SM}$ at some high-energy scale which we have not yet probed. %Finally, the new dynamics might be so strongly coupled that any particles charged under the new interactions confine to form neutral bound states with large binding energies, so that the additional interactions are not noticed at lower energies. 
In this paper we consider an alternative in which the new dynamics are so strongly coupled that particles charged under the new interactions confine to form neutral bound states, with %large binding energies, so that the additional interactions are not noticed at lower energies.
binding energies at the \TeV\ scale or larger.
%In this paper, 

We focus on a particular class of $\nn=1$ supersymmetric (\susy) gauge theories with product gauge groups of the form $SU(N)_1 \times SU(N)_2 \times \ldots \times SU(N)_k$. Our model includes one antisymmetric tensor $A_{\alpha\beta}$ and four quark fields $Q^{i}_\alpha$ charged under $SU(N)_1$, and a series of bifundamental fields $(\P_i)^\alpha_\beta$ charged under adjacent gauge groups $SU(N)_{i}\times SU(N)_{i+1}$ as shown in Table~\ref{table:UVmatter}. This theory is an extension of a model, $SU(N): (\yA + 4 \yF + N \yFb)$, which has been shown to confine~\cite{Berkooz:1995km,Pouliot:1995me,Poppitz:1996vh}.

\begin{table}[h]
\centering
%\begin{tabular}{| c | c | c  c c c c | c |} \hline
%   & $SU(4)_L$	& $G_1$ 	& $G_2$	& $G_3$	& $\ldots$ & $G_k$	& $SU(N)_R$		\\ \hline
%$Q$	& \yF		&	\yF	&		&		&		&		&		\\
%$A$	&		& \yA	&		&		&		&		&				\\
%$\P_1$&		& \yFb	&	\yF	&		&		&		&				\\
%$\P_2$&		& 		&	\yFb	&	\yF	&		&		&				\\ 
%$\vdots$ &	&		&		&	\yFb	& $\ddots$&	\yF	&		\\ 
%$\P_k$&		&		&		& 		&		&	\yFb	&	\yF			\\ \hline
%\end{tabular}
\includegraphics[scale=1.0]{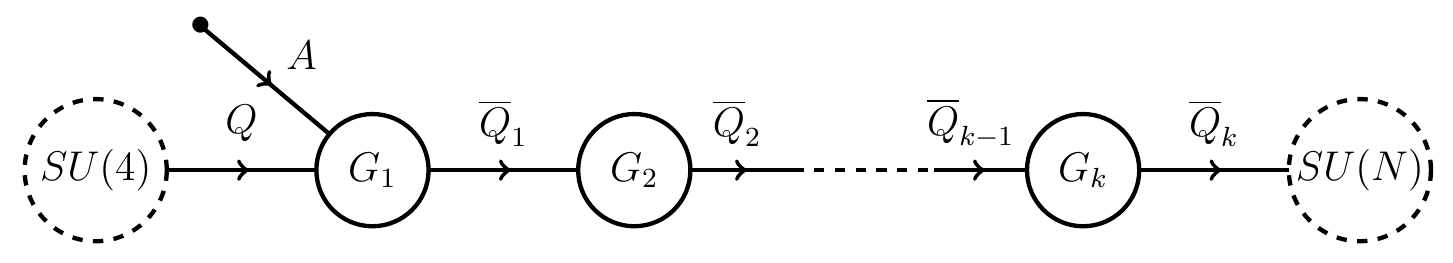}
\caption{The matter content of the proposed s-confining theory is shown as a moose diagram. Each $G_i$ represents a gauged $SU(N)$ group, while the dashed circles represent the $SU(4)_L \times SU(N)_R$ family symmetry.
}\label{table:UVmatter}
\end{table}

We propose in the language of \cite{Csaki:1996sm} that this $SU(N)^k$ model is ``s-confining:" that is, the theory confines smoothly in the infrared without breaking chiral symmetry, and it generates a non-vanishing superpotential that describes the interactions between the gauge invariant composite fields. 
Although the $\nn =1$ s-confining theories with a simple gauge group are fully classified~\cite{Csaki:1996zb}, our model is to the best of our knowledge the first with a nontrivial product gauge group and no superpotential.\footnote{An example based on $SU(5)\times SU(4)$ with a tree-level superpotential is discussed in~\cite{Hirayama:1998hu}.}

This $SU(N)^k$ product group model has two distinctive features which may be useful for model-building. First, there are no small gauge-singlet operators: the number of fields contained in every gauge invariant operator depends on $k$ or $N$. Second, the various $SU(N)_i$ subgroups generally confine at different scales $\Lambda_i$, with hierarchies based on the coupling constants $g_i$.
%This theory also shares many of the characteristics of confining theories that have already been studied. 

%Product groups of this form also appear when studying five-dimensional \susy\ gauge theories~\cite{Csaki:2001zx,Csaki:2001em,Skiba:2002nx}. If the fifth dimension is compact, then the low-energy behavior can be described by a Kaluza-Klein decomposition, in which the 5d gluon field is described as a tower of 4d Kaluza-Klein gluons with increasing masses. Each corresponds to a different broken $SU(N)$ gauge group. As the theory is probed at energies above the $k^\text{th}$ Kaluza-Klein mass, the effective gauge group is $SU(N)^k$.
%In this way, the theory shown in Table~\ref{table:UVmatter} can be thought of as a $k$ site deconstruction of a 5d \susy\ $SU(N)$ gauge theory with a $\ZZ_2$ orbifold. %with vectorlike fields $q$ and $\overline{q}$ in the bulk, and chiral fields $A$ and $Q$ on a 4d brane. 
Product groups of this form appear in studies of five-dimensional gauge theories~\cite{Csaki:2001em,Csaki:2001zx,Skiba:2002nx,Chang:2002qt}. The model shown in Table~\ref{table:UVmatter} can be interpreted as a $k$-site deconstruction of a 5d \susy\ $SU(N)$ gauge theory with a $\ZZ_2$ orbifold.
In the 5d theory the chiral fields $\{A, Q\}$ and $\barQ_k$ exist on opposing 4d branes, while the bifundamental $\barQ_i$ superfields correspond to a single bulk $\barQ$ field.
A natural hierarchy between the $\Lambda_i$ arises if the extra dimension is warped: for example, the model with $\Lambda_1> \ldots > \Lambda_k$ has $A$ and $Q$ on the ultraviolet brane and $\barQ_k$ on the infrared brane.

%In Section~\ref{sec:review} we review aspects of confinement in \susy\ gauge theories. In Section~\ref{sec:antiprops} we derive the coefficients in the dynamically generated superpotential for the s-confining theory with one antisymmetric tensor, which do not appear in the literature.
%Section~\ref{sec:main} addresses the question of whether the product group extension of this theory is also s-confining.
In Sections~\ref{sec:review} and \ref{sec:review2} we review the basic aspects of confining \susy\ gauge theories. In Section~\ref{sec:antiprops} we discuss more specific properties of the $A+4Q + N \barQ$ model with an $SU(N)$ gauge group, including the coefficients in its dynamically generated superpotential. These coefficients do not appear in the literature, so we include our derivation in Appendix~\ref{sec:appA}. Section~\ref{sec:main} contains a detailed discussion of the $SU(N)^k$ product gauge group models and our primary results. In Section~\ref{sec:misc} we suggest other product group models which may be s-confining, as well as several counter-examples.

%%%%%%%%
%Notes from Yuri:
%The review section isn't entirely necessary. If it's too long, people may be turned off. Review sections are mostly for the author, but can be useful (when applying to postdocs, etc). General message: make it shorter.
%The fermions are *combinations* of KK modes
%The superpotential IS generated: not "can be." In ADS it's generated by a calculable instanton effect: for s-confining it's a reflection of the classical constraints between the operators, which are also the quantum constraints.
%Make it clearer which things are introduction, and which topics are new.
%Reference for A+4Q+N\barQ is s-confining
%More explicitly define what "s-confinement" is.
%%%%%%%%

\subsection{Review: Seiberg Dualities} \label{sec:review}

It is generally difficult to analyze the infrared behavior of strongly coupled theories, due to the failure of perturbation theory in this limit. 
Seiberg, Intriligator and others have made this problem more tractable by exploiting some of the remarkable properties of supersymmetry, allowing some infrared properties of
\susy\ gauge theories to be calculated exactly \cite{Seiberg:1994pq,Intriligator:1994sm}.
Seiberg's infrared dualities between different phases of gauge theories were central to these developments. We summarize some of the results in this section; a more detailed review is given in \cite{Seiberg:1997vw}.

Seiberg found that in $SU(N)$ gauge groups with $F$ flavors of quarks and antiquarks, also known as \susy\ QCD, the infrared behavior of the $F=N$ and $F=N+1$ cases can be completely described by a set of gauge invariant operators, $M=Q\barQ$, $B = Q^N$, and $\barB = \barQ^N$. This dual theory has no gauge interactions, so the $F=N$ and $F= N+1$ theories are said to confine: every test charge can be ``screened" by creating quark-antiquark pairs from the vacuum, and a gauge-invariant Wilson loop obeys a perimeter law. 

%For $F>N+1$, the dual theory has a gauged $SU(F-N)$ symmetry, corresponding to an electric-magnetic duality. %For \susy\ QCD, the matter fields $Q$ and $\barQ$ transform as $\yF$ and $\yFb$, respectively; in the next section, we will add a field $A$ transforming in the $\yA$ representation
%This confined phase of the theory is conjectured to be dual to a ``Higgs phase," in which all fields have expectation values $\ev{\phi} \gg \Lambda$, and the gauge group is completely broken.
%These results spurred great interest in dualities for other \susy\ gauge theories, including models with fields in the antisymmetric $(\yA)$ representation. %+ F Q + (N-4+F) \barQ$. %It has been shown that such theories exhibit confinement if $F=3$ or $F=4$.

Classically, the gauge invariant operators obey particular constraints, following from the Bose symmetry of the superfields and the definitions of $M$, $B$, and $\barB$. For $F=N+1$,
\begin{eqnarray}
B_i M^i_j  &=& 0 \nonumber \\ 
M^i_j \barB^j &=& 0  \label{eq:fn1classical} \\
(M^i_j)^{-1} \det M &=& B_i \barB^j \,, \nonumber
\end{eqnarray}
while for $F=N$
\begin{equation}
\det M - B \barB = 0, \label{eq:fnclassical}
\end{equation}
where the indices $i$ and $j$ refer to the family $SU(F)$ symmetries of the $Q$ and $\barQ$.
It has been shown \cite{Davis:1983mz,Affleck:1984xz,Seiberg:1994bz} that \eqref{eq:fnclassical} is modified quantum mechanically:
\begin{equation}
\det M - B \barB = \Lambda^b, \label{eq:fnquantum}
\end{equation}
where $\Lambda^b$ is the holomorphic scale
\begin{equation}
\Lambda^b = \mu^b \exp\left\{-8\pi^2/g^2 + i \theta_\text{YM}  \right\} . \label{eq:holoscale}
\end{equation}
Here $\theta_\text{YM}$ is the $CP$-violating $\theta$-term of the $SU(N)$ gauge group, $g$ is the gauge coupling, and $b=3N-F=2N$ is derived from the $\beta$ function for the gauge coupling. 
The quantum-modified constraint \eqref{eq:fnquantum} can be enforced by a superpotential 
\begin{equation}
W = \lambda \left( \det M - B \barB - \Lambda^{2N} \right)
\end{equation}
if we introduce a Lagrange multiplier superfield $\lambda$.
At the origin of the classical moduli space, $M=B=\barB=0$, the UV family symmetry $SU(F)_L \times SU(F)_R \times U(1)_B$ is conserved. However, this point is not on the quantum-deformed moduli space given by \eqref{eq:fnquantum}, so the chiral symmetry is broken in the vacuum.

%\paragraph{Moduli space and infrared operators:}
%
%In a more general \susy\ gauge theory, the K\"ahler potential induces a ``$D$-term" potential involving the colored scalar fields, $V=\frac{1}{2} D^a D^a$, where
%\begin{equation}
%D^a = - g \sum_i \left( \phi_i^\star T^a_{r_\phi} \phi_i \right),
%\end{equation}
%for all scalar fields $\phi_i$ transforming in the representation $r_\phi$, where the $T^a_{r_\phi}$ are the generators of the gauge group. 
%%
%There is typically a continuous space of inequivalent vacua satisfying $V=0$, which is referred to as the moduli space. The degenerate vacua can be parametrized by holomorphic gauge-invariant operators~\cite{Luty:1995sd}. In the presence of a nonzero superpotential $W$, some of the flat directions may be lifted.

\subsection{Review: S-Confinement} \label{sec:review2}

In the $F=N+1$ case, the classical constraint equations are not modified. Instead, they are enforced by a dynamically generated superpotential~\cite{Intriligator:1995au}.
\begin{equation}
W_d = \frac{1}{\Lambda^{2N-1}} \left[ B M \barB -  \det M \right], \label{eq:wdsusyqcd}
\end{equation}
which has $\ev{M}=\ev{B}=\ev{\barB}=0$ as a solution to the equations of motion. This vacuum corresponds to confinement without chiral symmetry breaking, which we refer to as s-confinement. More precisely,
a theory is s-confining if~\cite{Csaki:1996sm}:
\begin{itemize}
\item All infrared degrees of freedom are gauge invariant composite fields;
\item The infrared physics is described by a smooth effective theory, which is valid everywhere on the moduli space (including the origin);
\item There is a dynamically generated superpotential.
\end{itemize}
For the effective theory to be smooth, there should be no gauge invariant order parameter that can distinguish the Higgs and confined phases of the theory. The infrared degrees of freedom must also satisfy the anomaly matching conditions. 

%As an example, consider \susy\ QCD with $F=N+1$. A dynamically-generated superpotential $W_d$ reproduces the classical constraints~\cite{Intriligator:1995au}:
%\begin{equation}
%W_d = \frac{1}{\Lambda^{2N-1}} \left[ B M \barB -  \det M \right], \label{eq:wdsusyqcd}
%\end{equation}
%and the $\ev{M}=\ev{B}=\ev{\barB}=0$ vacuum with unbroken chiral symmetry remains on the moduli space. 

Generally, the dynamically generated superpotential is determined up to an overall factor based on symmetry arguments, and by matching its equations of motion to the classical constraints. Its dependence on the holomorphic scale $\Lambda^b$ can be found either on dimensional grounds, or by requiring that $W_d$ is neutral under the anomalous $U(1)$ symmetry. %To find the overall factor in \eqref{eq:wdsusyqcd}, one can add large masses to two flavors of $Q\barQ$. Below this mass threshold, the infrared theory resembles $F=N-1$ \susy\ QCD, with a superpotential that is known exactly~\cite{Affleck:1984xz}.
%Other methods must be used to find the overall constant. For \susy\ QCD, the superpotential of the $F=N-1$ theory is known exactly~\cite{Affleck:1984xz}, so the factor in \eqref{eq:wdsusyqcd} can be derived by integrating out two quark flavors from the $F=N+1$ theory. 

The requirement that a superpotential is dynamically generated adds a powerful constraint on the matter content of any s-confining theory. 
An $\nn=1$ \susy\ theory with $f$ massless matter superfields has a classical family symmetry of rank $f+1$ including the $R$ symmetry, but the $G^2 U(1)$ anomaly removes one linear combination of the $U(1)$ family symmetries. %so that the rank of the family symmetry matches the number of matter fields.
This allows us to define a $U(1)_R$ symmetry such that exactly one of the matter superfields $\phi_i$ has $R$ charge, $q_i$, with all other fields neutral. Using the normalization in which the gauginos have $R$ charge $+1$, cancellation of the $G^2 U(1)_R$ anomaly requires that
%\begin{equation}
%(q_i - 1) \mu_i + \sum_{j\neq i} (-1)\mu_j + (+1) \mu_G = 0,
%\end{equation}
%or 
\begin{equation}
q_i = \frac{1}{\mu_i} \left[ \sum_{j} \mu_j - \mu_G \right], \label{eq:Rchargei}
\end{equation}
where $\mu_j$ and $\mu_G$ are the Dynkin indices of the matter fields $\phi_j$ and the gluinos, respectively, with the normalization $\mu(\yF) =1$. 
For the dynamically generated superpotential to have $R$ charge $+2$ under any of the possible anomaly-free $R$ symmetries, it must have the form
\begin{equation}
W \sim \prod_{i} \left[ \phi_i^{2/q_i} \right] = \prod_i \left( \phi_i^{\mu_i} \right)^{2/\left[ \sum_j \mu_j - \mu_G \right]}.
\end{equation}
The matter content must therefore satisfy the index constraint of Csaki {\it et al.}~\cite{Csaki:1996sm}:
\begin{equation}
\sum_j \mu_j - \mu_G = 2 . \label{eq:indexconst}
\end{equation}
In~\cite{Csaki:1996zb} this index constraint is used to find all $\mathcal N=1$ s-confining theories with one gauge group and no tree-level superpotential.
%The index constraint reduces the number of possible s-confining theories to a countable set, including
Both $F=N+1$ \susy\ QCD and the $A+4Q + N\barQ$ model are included. 

In theories with a product gauge group this constraint is relaxed: the number of fields exceeds the rank of the family symmetry, and it is no longer possible to identify a unique $R$ symmetry for each field.
%The index combination can also be used to identify theories with quantum-deformed constraints ($\sum_j \mu_j - \mu_G = 0$) and theories with a superpotential that can be generated by a one-instanton effect ($\sum_j \mu_j - \mu_G = -2$).
% is also useful for identifying theories with quantum-deformed constraints: for these theories,
%\begin{equation}
%\sum_j \mu_j - \mu_G = 0 .  \label{eq:QDindexconst}
%\end{equation}
%Grinstein and Nolte \cite{Grinstein:1997zv} enumerate the theories which satisfy this constraint.
%The Seiberg-Witten theories with Coulomb phases have also been catalogued \cite{Csaki:1997zg,Csaki:1998dp}.
%Similarly, theories with $\sum_j \mu_j - \mu(G) = -2$ have the right quantum numbers for a superpotential to be generated by one-instanton effects.

%\paragraph{Product groups:}
%Product gauge groups are generally less well-studied, with a few notable exceptions.
%In Intriligator's ``integrating in" technique~\cite{Intriligator:1994uk}, new dualities are found by treating a UV field as a bound state of a new strongly coupled gauge group.
%Several dualities have been found for product groups using these techniques~\cite{Poppitz:1996vh,Hirayama:1998hu,Csaki:2001zx}.
%
%As discussed previously, deconstruction of 5d gauge theories 
%Several dualities have been found for product groups using these techniques~\cite{Poppitz:1996vh,Hirayama:1998hu,Csaki:2001zx}.

\subsection{$SU(N)$ with antisymmetric tensor} \label{sec:antiprops}

Properties of the $\yA + F\, \yF + (N+F-4)\ \yFb\ $ model have been studied by several authors \cite{Berkooz:1995km,Poppitz:1995fh,Pouliot:1995me,Terning:1997jj}. In the $F=2$ case there is a superpotential generated by a one-instanton effect; for $F=3$ the theory confines, with a quantum-deformed moduli space that induces dynamical symmetry breaking; and for $F=4$, the theory is s-confining. 
The quantum modified constraints have been derived in~\cite{Poppitz:1995fh} for $F=3$, but the classical constraints for the $A + 4 Q + N \barQ$ model do not appear in the literature. %In this section we calculate these missing elements, specifically the coefficients that appear in the dynamically generated superpotential.
We derive the relative coefficients of the dynamically generated superpotential in Appendix~\ref{sec:appA}, and quote the results in this section.

%The number of independent IR degrees of freedom is equal to the dimension of the moduli space, $\dim M_0$. This is also equal to the number of UV fields, modulo the gauge degrees of freedom:
%\begin{equation}
%\dim M_0 = \left( \frac{N (N-1)}{2} + 4 N + N^2 \right) - (N^2-1) = \frac{N (N-1)}{2} + 4 N + 1 .
%\end{equation}
%If the number of IR operators $N_\text{ops}$ exceeds $\dim M_0$, then there must be some number of constraint equations $N_\text{con}$ such that
%\begin{equation}
%N_\text{con} =N_\text{ops} - \dim M_0.
%\end{equation}

%The $D$ flat directions can be described in a non-holomorphic way by considering solutions to $D^a D^a = 0$, where
%\begin{equation}
%D^a = - g \left( Q^{\star\alpha}_i (T^a_{\ysub{\yF}})^\beta_\alpha Q^i_\beta + \barQ^{\star j}_\alpha (T^a_{\ysub{\yFb}})_\beta^\alpha \barQ_j^\alpha + A^{\star \beta\alpha} (T^a_{\ysub{\yA}})_{\alpha\beta}^{\delta\epsilon} A_{\delta\epsilon} \right), \label{eq:Dflatness}
%\end{equation}
%which can be rearranged to extract the constraint
%\begin{equation}
%Q^{\star\alpha}_i Q^i_\beta  - \barQ_j^\alpha \barQ^{\star j}_\beta + 2 A^{\star \alpha\gamma} A_{\gamma\beta } = \rho \delta^\alpha_\beta  \label{eq:dflatness2}
%\end{equation}
%for some constant $\rho$. We derive this relationship in Appendix~\ref{sec:dflatness}.
%In Appendix~\ref{sec:dflatness} we derive this relationship, and discuss particular solutions of \eqref{eq:dflatness2}. 

\paragraph{Infrared operators:}
In the $A+4 Q + N\barQ$ model, the set of gauge invariant operators changes based on whether $N$ is even or odd. This is due to the $\yA$ representation:
if $N=2m$ is even, then the gauge invariants include the antisymmetrized products $(A^m)$, $(A^{m-1} Q^2)$, and $(A^{m-2} Q^4)$, while for odd $N=2m+1$ the gauge invariants include $(A^m Q)$ and $(A^{m-1} Q^3)$. %Other invariants can be formed by replacing $A_{\alpha\beta}$ with $(Q^i_\alpha Q^j_\beta -Q^i_\beta Q^j_\alpha)$. 
%For $A + F Q + (N+F-4) \barQ$ with $N\geq F$, the series in $Q^p$ terminates %at $Q^{F}$: that is, 
%at $Q^{F+1} = 0$ for antisymmetric products.

Below, we define the simplest gauge invariant operators for the $N=2m$ and $N=2m+1$ models. Both cases include the operators $(Q\barQ)$, $(A\barQ^2)$, and $(\barQ^N)$:
\begin{eqnarray}
J^i_j &=& Q^i_\alpha \barQ^\alpha_j , \label{eq:defJ} \\
K_{j_1 j_2} &=& A_{\alpha\beta} \barQ^{\alpha}_{j_1} \barQ^\beta_{j_2} , \\
Z &=& \det \barQ \;=\; \frac{\epsilon_{\alpha_1 \ldots \alpha_N} \epsilon^{j_1 \ldots j_N} }{N!} \left( \barQ_{j_1}^{\alpha_1}  \barQ_{j_2}^{\alpha_2}  \ldots  \barQ_{j_N}^{\alpha_N} \right) .
\end{eqnarray}
For even $N\geq 4$, we also add the gauge invariants
\begin{eqnarray}
U &=& \Pf A \;=\; \frac{\epsilon^{a_1 a_2 \ldots a_N}}{2^m m!} \left( A_{a_1 a_2} A_{a_3 a_4} \ldots A_{a_{N-1} a_N} \right)  , \label{eq:defU}\\
V_{i_1 i_2} &=& \frac{\epsilon^{a_1 a_2 \ldots a_N}}{2^{m-1} (m-1)! 2!} \left( A_{a_1 a_2} A_{a_3 a_4} \ldots A_{a_{N-3} a_{N-2} } \right) Q_{a_{N-1}}^{i_1} Q_{a_N}^{i_2}   , \\
\ww &=& \frac{\epsilon^{a_1 a_2 \ldots a_N}}{2^{m-2} (m-2)!} \frac{\epsilon_{j_1 j_2 j_3 j_4} }{4!} \left( A_{a_1 a_2} A_{a_3 a_4} \ldots A_{a_{N-5} a_{N-4} } \right) Q_{a_{N-3}}^{j_1} Q_{a_{N-2}}^{j_2} Q_{a_{N-1}}^{j_3} Q_{a_N}^{j_4} ,
\end{eqnarray}
whereas for odd $N \geq 5$ we include 
\begin{eqnarray}
X^j &=& \frac{\epsilon^{a_1 a_2 \ldots a_N}}{2^m m!} \left( A_{a_1 a_2} A_{a_3 a_4} \ldots A_{a_{N-2} a_{N-1}} \right) Q^j_{a_N} , \\
Y_j &=& \frac{\epsilon^{a_1 a_2 \ldots a_N}}{2^{m-1} (m-1)!} \frac{\epsilon_{j  j_2 j_3 j_4}}{3!} \left( A_{a_1 a_2} A_{a_3 a_4} \ldots A_{a_{N-4} a_{N-3} } \right) Q_{a_{N-2}}^{j_2} Q_{a_{N-1}}^{j_3} Q_{a_N}^{j_4} . \label{eq:defY}
\end{eqnarray}
The numeric coefficients absorb the combinatoric factors from the $\epsilon$ tensors, with the convention $\epsilon_{123 \ldots N} = +1$.
In general, we reserve the indices $a, b,\alpha, \beta$ for gauge groups, and use the indices $i, j$ to refer to family symmetries.
Superscripts and subscripts are chosen for visual clarity, and do not signify any particular group representation.

It is useful to classify the $\{U, V, \ww, X, Y, Z\}$ fields as ``baryons" and the $J$ and $K$ fields as ``mesons," to separate the operators which scale with $N$ from those which are independent of $N$.   
The transformation properties of these operators under the family symmetries are shown in Table~\ref{fig:evenoddtable}.
There is a continuous family of equivalent $U(1)_A  \times U(1)_B \times U(1)_R$ charge assignments, but the choice shown in Table~\ref{fig:evenoddtable} is particularly convenient.
%It is also useful to keep track of the anomalous $U(1)$ symmetry: under this spurious symmetry, the charge of the holomorphic scale $\Lambda$ is determined by the anomaly coefficient for $G^2 U(1)$.

\begin{table}
\begin{center}
\begin{tabular}{| c | c | c c | c c  c | c |} \hline
	& $G$ 	& $SU(4)_L$& $SU(N)_R$& $U_A$	& $U_B$	& $U_R$	& $U_1$		\\ \hline
$A$	& \yA	&		&			&	$-4$	& $-1$	&	0	&	0		\\
$Q$	& \yF		&	\yF	&			& $N-2$	& $-1/2$	& $1/2$	&	0		\\
$\barQ$& \yFb	&		&	\yF		& $0$	& $1$	& 	0	&	1		\\ \hline
$\Lambda^b$	& 	&	&			& $0$	& $0$	& 	0	& $N$		\\ \hline\hline
$J$	&		&	\yF	&	\yF		& $N-2$	& $1/2$	& $1/2$	&	1		\\
$K$	&		&		&	\yA		& $-4$	& $1$	&	0	&	2	\\ \hline 
$Z$	&		&		&			& $0$	& $N$	&	0	&	$N$	\\ \hline\hline
$U$	&		&		&			& $-2N$	&$-N/2$	& 	0	&	$0$	\\
$V$	&		&	\yA	&			& $0$	&$-N/2$	&	1	&	$0$ \\
$\ww$&		&		&			& $2N$	&$-N/2$	&	2	&	$0$ \\ \hline
$X$	&		&	\yF	&			& $-N$	&$-N/2$	&  $1/2$	&	$0$	\\
$Y$	&		&	\yFb	&			& $N$	&$-N/2$	& $3/2$	&	$0$ \\ \hline 
\end{tabular}
\end{center}
\vspace{-0.5cm}
\caption{The transformation properties of the UV and IR fields under the family $SU(4)_L \times SU(N)_R \times U(1)_A \times U(1)_B \times U(1)_R$ symmetry for the $F=4$ model are shown, along with the charges under the spurious $U(1)_1$. The operators $J$, $K$, and $Z$ are defined whether $N$ is even or odd; the fields $U$, $V$ and $\ww$ are specific to the even $N$ case, while the fields $X$ and $Y$ correspond to the odd $N$ case. The $U(1)_R$ charges listed refer to the scalar component of each superfield. } \label{fig:evenoddtable}
\end{table}

For $N=4$, the theory contains four flavors of $Q + \barQ$. This value of $N$ is unique in that both $m_A \Pf A$ and $m^i_j Q_i^\alpha \barQ_\alpha^j$ are gauge-invariant mass terms: if these masses are large compared to $\Lambda$, then every field can be integrated out above the confinement scale. This special case is discussed in Section~\ref{sec:su4LR}.
For $N=3$ the $\yA$ and $\yFb$ representations are equivalent, and the $A + 4 Q + 3 \barQ$ model reduces to \susy\ QCD with $F=4$.

As discussed in Section~\ref{sec:review}, the form of the dynamically generated superpotential is determined by the representations of the matter fields. For the $A + 4 Q + N \barQ $ model,
\begin{equation}
W_d \sim \sum  \frac{A^{N-2} Q^4 \barQ^N}{\Lambda^b} .
\end{equation}
The sum includes all possible gauge-invariant contractions of the group indices, with some relative coefficients:
%This $W_d$ has $R$ charge $+2$, and it is invariant under the family $SU(4)_L \times SU(N)_R \times U(1)_A \times U(1)_B$ symmetry as well as the spurious $U(1)$.
%The superpotential can include any of the following terms: 
\begin{eqnarray}
W_\text{odd N} &\sim& \frac{1}{\Lambda^b} \left[ X Y Z + X K^{m-1} J^3 + Y K^m J \right] , \label{eq:wpodd} \\
W_\text{even N} &\sim& \frac{1}{\Lambda^b} \left[U \ww Z + V^2 Z + U K^{m-2} J^4 + V K^{m-1} J^2 + \ww K^m \right]. \label{eq:wpeven}
\end{eqnarray}
%Each term in $W_d$ has a coefficient which is determined by requiring that the equations of motion reproduce the classical constraints.
Both $\ff_\text{odd} = \{J, K, X, Y, Z\}$ and $\ff_\text{even} = \{J, K, U, V, \ww, Z\}$ satisfy the t'~Hooft anomaly matching conditions for the mixed $SU(4)^2 U(1)$ and $SU(N)^2 U(1)$ anomalies, the various $U(1)^3$ anomalies, and the mixed $U(1)$ gravitational anomalies, for all $U(1)$ symmetries listed in Table~\ref{fig:evenoddtable} except for  $U(1)_1$. The $G_1^2 U(1)_1$ anomaly breaks $U(1)_1$ explicitly at the scale $\Lambda_1$, so it is not a symmetry of the infrared theory.

\paragraph{Dynamically generated superpotential:}

The number of infrared operators, $\dim \ff$, is larger than the dimension of the moduli space, $\dim M_0 = N(N-1)/2 + 4N + 1$. For $N=2m+1$,
\begin{eqnarray}
\dim \{ J, K, X, Y, Z  \} &=& \left( 4N + \frac{N(N-1)}{2}  + 4 + 4 + 1 \right),   %\nonumber\\ 
%N_\text{con} &=&  8, \label{eq:nconstr}
\end{eqnarray}
and for $N=2m$,
\begin{eqnarray}
\dim \{ J, K, U, V, \ww, Z  \}  &=& \left( 4N + \frac{N(N-1)}{2} + 1 + \frac{4(3)}{2} + 1 + 1 \right),  %\nonumber\\
%N_\text{con} &=&  8 .
\end{eqnarray}
implying for both cases that the number of constraints is
\begin{equation}
N_\text{con} = \dim \ff - \dim M_0 = 8. \label{eq:nconstr}
\end{equation}
%To find these constraints, we look for products of infrared operators that have the same form in terms of $(A^r Q^s \barQ^t)$: for example, $XZ \sim (A^m Q \barQ^N)\sim K^m J$. In the classical limit, this relationship between $XZ$ and $K^m J$ should become exact, with some coefficient that we can calculate.

For odd $N$, the eight constraints are
%To derive the constraint equations, we consider a generic point on the moduli space with $\ev{\phi_i} \gg \Lambda$ for all gauge invariant operators $\phi_i$. In this classical limit the description of $\phi_i$ as a product of $A$, $Q$ and $\barQ$ becomes exact, producing the relationships 
%For odd $N$, we find that
\begin{eqnarray}
X^i Z &=& \frac{\epsilon^{j_1 j_2 \ldots j_N}}{2^m m!}  \left(K_{j_1 j_2} K_{j_3 j_4} \ldots K_{j_{N-2} j_{N-1}} \right) J^i_{j_N} \nonumber \\
Y_i Z &=& \frac{\epsilon^{j_1 j_2 \ldots j_N} \epsilon_{i i_2 i_3 i_4} }{2^{m-1} (m-1)! 3! } \left(K_{j_1 j_2} K_{j_3 j_4} \ldots K_{j_{N-4} j_{N-3} }\right) J^{i_2}_{j_{N-2}} J^{i_3}_{j_{N-1}} J^{i_4}_{j_{N}} \, , \label{eq:classconstodd}
\end{eqnarray}
%for odd $N$, and
%where $i=1\ldots 4$ runs over the $SU(4)$ family index.
%If instead $N$ is even, the eight constraints are
while for even $N$
\begin{eqnarray}
U Z &=& \frac{\epsilon_{j_1 \ldots j_N} }{2^m m!} K_{j_1 j_2} K_{j_3 j_4} \ldots K_{j_{N-1} j_N}  \;=\; \Pf K , \nonumber\\
V_{i_1 i_2} Z &=& \frac{\epsilon_{j_1 \ldots j_N} }{2^{m-1} (m-1)!} \frac{\epsilon_{i_1 i_2 i_3 i_4} }{2!} K_{j_1 j_2} K_{j_3 j_4}  \ldots K_{j_{N-3} j_{N-2}} J_{j_{N-1}}^{i_3} J_{j_{N}}^{i_4} , \label{eq:classconsteven} \\
\ww Z &=& \frac{\epsilon_{j_1 \ldots j_N} }{2^{m-2} (m-2)!} \frac{\epsilon_{i_1 i_2 i_3 i_4} }{4!} K_{j_1 j_2} K_{j_3 j_4}  \ldots K_{j_{N-5} j_{N-4}} J_{j_{N-3}}^{i_1} J_{j_{N-2}}^{i_2} J_{j_{N-1}}^{i_3} J_{j_{N}}^{i_4}. \nonumber
\end{eqnarray}
The index $i=1\ldots4$ refers to the $SU(4)$ family symmetry.

%These are not the only classical relationships we can write down: for example, the fields also obey
%\begin{equation}
%X^i Y_i = 0
%\end{equation}
%or
%\begin{equation}
%U \ww = \frac{\epsilon^{i_1 i_2 i_3 i_4}  }{ 2^2 2! }V_{i_1 i_2} V_{i_3 i_4},
%\end{equation}
%which follow from Eqs.~(\ref{eq:classconstodd}) and~(\ref{eq:classconsteven}), respectively. \rev{Tangential?}
%
%In general, the constraints can be found in the classical limit, far away from the origin along a $D$-flat direction, by rearranging the gauge index contractions in products of the infrared operators.
%We use this method in Appendix~\ref{sec:appA} to find the coefficients in Eqs.~(\ref{eq:classconstodd}) and~(\ref{eq:classconsteven}).

By taking partial derivatives of \eqref{eq:wpodd} and \eqref{eq:wpeven} and matching the equations of motion to the classical constraints, %it becomes clear that each classical constraint appears as one of the equations of motion, $\partial W/\partial \phi$. After some work, 
one can determine the relative coefficient of each term in the dynamically generated superpotential. The results appear below:
\begin{eqnarray}
W_\text{odd} &=& \frac{\alpha}{\Lambda^b} \bigg\{ X^i Y_i Z - \frac{\epsilon^{j_1 \ldots j_N} \epsilon_{i_1 \ldots i_4} }{ 2^{m-1} (m-1)! 3!} X^{i_1} (K_{j_1 j_2 } \ldots K_{j_{N-4} j_{N-3} } )J^{i_2}_{j_{N-2}} J^{i_3}_{j_{N-1}} J^{i_4}_{j_{N}} \nonumber\\&&\
- \frac{\epsilon^{j_1 \ldots j_N}  }{ 2^{m} m!} Y_i (K_{j_1 j_2} \ldots K_{j_{N-2} j_{N-1}} ) J_{j_N}^i \bigg\} ; \label{eq:w1odd} \\
W_\text{even} &=& \frac{\alpha}{\Lambda^b} \bigg\{ U \ww Z - \frac{\epsilon_{i_1 \ldots i_4} }{2^2 2!} V^{i_1 i_2} V^{i_3 i_4} Z - \ww\ \Pf K  \nonumber\\&&\
- \frac{\epsilon_{j_1 \ldots j_N} }{2^{m-2} (m-2)!} \frac{\epsilon_{i_1 i_2 i_3 i_4} }{4!} U ( K_{j_1 j_2} \ldots K_{j_{N-5} j_{N-4}}) (J_{j_{N-3}}^{i_1} \ldots J_{j_{N}}^{i_4}) \nonumber\\&&\
+  \frac{\epsilon_{j_1 \ldots j_N} \epsilon_{i_1 i_2 i_3 i_4}}{4\cdot 2^{m-1} (m-1)! }   V^{i_1 i_2} (K_{j_1 j_2} \ldots K_{j_{N-3} j_{N-2}}) J^{i_3}_{j_{N-1}} J^{i_4}_{j_N} 
\bigg\} . \label{eq:w1even}
\end{eqnarray}
As in \susy\ QCD, the overall factor $\alpha$ cannot be determined by symmetry arguments. In principle, it is possible to add heavy quark masses and integrate out two flavors of $(Q \barQ)$ so as to match the $F=2$ model, whose superpotential can be calculated from a one-instanton calculation analogous to $F=N-1$ \susy\ QCD. 
In our present study we do not perform this calculation. 

It is useful, however, to consider the phases of $\alpha$ and $\Lb{}$. As defined in \eqref{eq:holoscale}, the phase of $\Lb{}$ is determined by the $CP$-violating $\theta_\text{YM}$ parameter. The phase of $\alpha$ is also unknown: however, because $W_d$ is charged under an unbroken $U(1)_R$ symmetry, it can be rotated by a phase without affecting the Lagrangian $\mathcal L \sim \int d\theta^2 W$, so as to make $\alpha$ real.%, leaving the phase of $\Lb{}$ determined by the parameter $\theta_\text{YM}$. 

%It can be seen from Table~\ref{fig:evenoddtable} that discrete $U(1)_1$ rotations of $2\pi/N$ have no effect on the superpotentials $W_\text{odd}$ and $W_\text{even}$, even though $\barQ$ transforms nontrivially. This suggests that the anomalous $U(1)$ symmetry is broken to the discrete group $\ZZ_N$ in the infrared theory.

\section{Product Group Extension for an S-Confining Theory} \label{sec:main}

%Berkooz:1995km
%Intriligator:1994uk
%The work of Chang and Georgi~\cite{Chang:2002qt} on $SU(N)^k$ extensions to $F=N$ \susy\ QCD is most immediately relevant to our present study.  As in our model, Chang and Georgi add a series of bifundamentals charged under a gauged $SU(N) \times \ldots \times SU(N)$ group. They devote particular attention to the $N=2$ case, where it is possible to add gauge-invariant mass terms for any of the fields. We refer to some of their results in Section~\ref{sec:disorderedWd}. 

%Now that we have discussed the properties of the $A+4Q+N\barQ$ model, we are equipped to study the product group model proposed in Table~\ref{table:UVmatter}. 

Our interest in the product group model of Table~\ref{table:UVmatter} is motivated by an observation from the $G_1 \times G_2$ case, in which the family symmetry $G_2=SU(N)_R$ of the $\barQ$ is weakly gauged. %To cancel the $G_2$ anomaly we add $N$ fields $\P_2$, charged as $(1, \yFb)$ under $G_1 \times G_2$.
In the confined phase of $G_1$, there are three types of operators charged under $G_2$: one antisymmetric $K=\yA$, four quarks $J=\yF$, and $N$ antiquarks $\P_2 = \yFb$. Remarkably, this is identical to the original s-confining model.

The model described in Section~\ref{sec:antiprops} can be extended indefinitely by adding more gauge groups $G_i$ and bifundamental matter $\P_i$. As long as $\Lambda_1 > \Lambda_2 \ldots >\Lambda_i> \Lambda_{i+1}$, confinement under $G_i$ always produces mesons charged as $\yA + 4 \yF$ under $G_{i+1}$.
This is the model shown in Table~\ref{table:UVmatter}, where the gauge group is $G_1 \times \ldots \times G_k$.
In this section we devote our attention to the question: is this $SU(N)^k$ theory s-confining, or is s-confinement disrupted by the product group?

There are two obvious ways in which the $K+ 4J + N \P_2$ ``k=2" model differs from the original (``k=1") s-confining theory. First, in the $k=1$ model there is no tree-level superpotential, but in the $k=2$ case there is a superpotential from $G_1$ confinement that may alter how $\{K, J, P\}$ confine under $G_2$. Luckily, inspection of the classical constraints shows that $K$, $J$, and $\P_2$ may be varied freely, as long as the baryon products $\{ UZ, VZ, \ww Z\}$ or $\{XZ, YZ\}$ vary in accordance with Eqs.~(\ref{eq:classconstodd}) and~(\ref{eq:classconsteven}). 
The second main difference is that under $G_2$, the classical moduli space is modified quantum mechanically. For the $k\geq 2$ theory to be s-confining, we must determine whether or not the origin remains on the moduli space.

%In this section we determine the general form for the dynamically generated superpotential using symmetry arguments, and we find a set of gauge-invariant operators which satisfy the anomaly matching conditions. We consider how the classical constraints may receive quantum corrections, and if the origin of moduli space remains on the quantum moduli space.

Of the existing literature regarding \susy\ product groups, the work of Chang and Georgi~\cite{Chang:2002qt} on $SU(N)^k$ extensions to $F=N$ \susy\ QCD is particularly useful to our present study.  Our method also has some similarities to deconfinement~\cite{Intriligator:1994uk,Berkooz:1995km}, particularly in Section~\ref{sec:misc} when we consider $Sp(2N)$ groups. % however, rather than taking only the $\yA$ field to be a bound state under an $Sp(2m)$ group, the entire set $\{\yA+4\yF\}$ is taken to be composite.

\subsection{Infrared Operators} \label{sec:infraops}
To understand the infrared behavior of the theory, we develop in this section a basis of gauge invariant operators which describe the moduli space and obey anomaly matching conditions.
%In this section we develop a basis of gauge invariant operators which describe the moduli space, while also obeying the 't~Hooft anomaly matching conditions. 
Then in Sections~\ref{sec:orderedWd} and~\ref{sec:disorderedWd}, we find the dynamically generated superpotential and perform some consistency checks.

%For every $G_i$ there is a $U(1)_i$ symmetry with a nonzero anomaly coefficient, breaking the approximate UV symmetry to: 
%\begin{equation}
%SU(4)_L\times SU(N)_R \times U(1)_R \times U(1)^{k+2}   \longrightarrow  SU(4)_L \times SU(N)_R \times U(1)_R \times U(1)_A\times U(1)_B.
%\end{equation}
Let us define a basis for the anomalous $U(1)$ charges, $U(1)_{j=1\ldots k}$, such that the anomaly coefficient $\mathcal A(G_i^2 U(1)_j)$ is zero if and only if $i \neq j$, as shown in Table~\ref{table:UVsymmetry}. Each $U(1)_i$ is explicitly broken at a scale associated with $\Lambda_i$,
so that the approximate UV symmetry is broken to
\begin{equation}
SU(4)_L\times SU(N)_R \times U(1)_R \times U(1)^{k+2}   \longrightarrow  SU(4)_L \times SU(N)_R \times U(1)_R \times U(1)_A\times U(1)_B.
\end{equation}
The $U(1)_i$ charges of the $\Lb{i}$ are determined by the $G^2 U(1)$ anomaly coefficients. Note that $b=2N-1$ for $\Lambda_1^b$, while $b=2N$ for $\Lambda_{i\neq 1}^b$.

\begin{table}[h]
\centering
\begin{tabular}{| c | c  c  c c c | c  c || c c c | c c c c c |} \hline
   &  $G_1$ 	& $G_2$	& $G_3$	& $\ldots$ & $G_k$	&$SU(4)$& $SU(N)$	&$U_A$	& $U_B$	& $U_R$	&$U_1$	& $U_2$	&$U_3$	&$\ldots$	& $U_k$	\\ \hline
$Q$	&	\yF	&		&		&		&		& \yF		&		&$N-2$	& $-1/2$	& $1/2$	&	0	&	0	&	0	&		&	0	\\
$A$	& \yA	&		&		&		&		&		&		& $-4$	&	$-1$	&	0	&	0	&	0	&	0	&		&	0	\\
$\P_1$& \yFb	&	\yF	&		&		&		&		&		&	0	&	1	&	0	&	1	&	0	&	0	&		&	0	\\ \hline
$\P_2$& 		&	\yFb	&	\yF	&		&		&		&		&	0	&	$-1$	&	0	&	$-1$	&	1	&	0	&$\ldots$	&	0	\\
$\P_3$&		&		&	\yFb	& $$		&		&		&		&	0	&	1	&	0	&	1	&	$-1$	&	1	&		&	0	\\ 
$\vdots$ &	&		&		& $\ddots$&	\yF	&		&		&	0	&$\vdots$	&	0	&$\vdots$&		&$\vdots$	&		&	0	\\ 
$\P_k$&		&		& 		&		&	\yFb	&		&	\yF	&	0	&$\pm1$	&	0	&$\pm1$	&$\mp1$	&$\pm1$	& $\ldots$	&	1	\\ \hline\hline
$\Lambda_1^b$ &&		&		&		&		&		&		&	0	&	0	&	0	&	$N$	&	0	&	0	&		&	0	\\
$\Lambda_2^b$ &&		&		&		&		&		&		&	0	&	0	&	0	&	0	&	$N$	&	0	&		&	0	\\
$\Lambda_3^b$ &&		&		&		&		&		&		&	0	&	0	&	0	&	0	&	0	&	$N$	&		&	0	\\	
$\vdots$ &	&		&		&		&		&		&		&$\vdots$	&$\vdots$	&$\vdots$	&$\vdots$	&		&		& $\ddots$&	0	\\
$\Lambda_k^b$ &&		&		&		&		&		&		&	0	&	0	&	0	&	0	&	0	&	0	&		&	$N$	\\ \hline
\end{tabular}
\caption{Matter content of the proposed s-confining theory, showing the transformation properties under the gauged $SU(N)^k$ and the $SU(4)_L \times SU(N)_R\times U(1)_A \times U(1)_B \times U(1)_R$ family symmetry. The spurious $U(1)_{i=1\ldots k}$ charges are also shown. The alternating $(\pm)$ factors in the $\P_k$ charges depend on whether $k$ is odd or even: the upper choice corresponds to odd $k$.}  
\label{table:UVsymmetry}
\end{table}

From Table~\ref{table:UVsymmetry}, it is clear that combinations of the form
\begin{align*}
\left(\frac{\P_1^N \P_2^N}{\Lambda_2^b} \right) ,  \left(\frac{\P_2^N \P_3^N}{\Lambda_3^b} \right), \ldots  \left(\frac{\P_{k-1}^N \P_k^N}{\Lambda_k^b} \right)
\end{align*}
are neutral under all of the symmetries, including the spurious $U(1)_i$.
Therefore, the dynamically generated superpotential has the form
\begin{equation}
W_d \sim \sum_{p_2  \ldots p_k} \left\{ \left(\frac{A^{N-2} Q^4 \P_1^N }{\Lambda_1^b} \right)\left(\frac{\P_1^N \P_2^N}{\Lambda_2^b} \right)^{p_2} \left(\frac{\P_2^N \P_3^N}{\Lambda_3^b} \right)^{p_3} \ldots \left(\frac{\P_{k-1}^N \P_k^N}{\Lambda_k^b} \right)^{p_k} \right\} \label{eq:wdgeneral}
\end{equation}
for some powers $p_i = 0, 1, \ldots$ for each $i=2, 3, \ldots k$. Any such superpotential has an $R$ charge of $+2$ under all of the possible $U(1)_R$ symmetries.
Before we can find the individual terms that appear in $W_d$, it is necessary to understand the equations of motion between the infrared operators.

%A basis of infrared operators must include sufficiently many degrees of freedom to cover the moduli space. For the $G_1 \times\ldots\times G_k$ gauge group with fields $\{A, Q, \P_1, \ldots, \P_k\}$, the dimension of the moduli space is
%\begin{equation}
%\dim M_0(k) = \frac{N(N-1)}{2} + 4 N + k N^2 - k (N^2 - 1) = 4 N + \frac{N(N-1)}{2} + k . \label{eq:dimM0k}
%\end{equation}
To find a set of gauge invariant operators in the far infrared, let us consider the ordered case $\Lambda_1 \gg \Lambda_2 \gg \ldots \gg \Lambda_k$.
As discussed in Section~\ref{sec:antiprops}, $G_1$ confinement produces the operators
\begin{align}
J_1 = (Q \P_1) &,&
K_1 = (A \P_1^2) &,&
Z_1 = (\P_1^N),
\end{align}
\begin{align}
U_1 = (A^m) ,&&
V_1 = (A^{m-1} Q^2), &&
\ww_1 = (A^{m-2} Q^4) &;&
X_1 = (A^m Q), &&
Y_1 = (A^{m-1} Q^3), &&
\end{align}
where $J_1$ and $K_1$ are charged under $G_2$. Although $U(1)_1$ is broken, the $U(1)_2 \times \ldots \times U(1)_k$ symmetry is approximately preserved above the scale $\Lambda_2$, adding $\mathcal O(k^3)$ anomaly coefficients that must be calculated. 

This is the benefit of the strategically-defined $U(1)_i$ charges shown in Table~\ref{table:UVsymmetry}: the fields $\{Q, A, \P_1\}$ are neutral under $U(1)_2 \ldots U(1)_k$, and all of these anomaly matching conditions are trivially satisfied.
The fields $J_1$ and $K_1$ transform similarly to $Q$ and $A$ under the non-Abelian symmetries, but their $U(1)_B$ charges are different, as shown in Table~\ref{table:UVsymmetry2}.

\begin{table}[h]
\centering
\begin{tabular}{| c | c   c c c | c  c || c c c | c c c c |} \hline
   	&  $G_2$ 	& $G_3$	& $\ldots$ & $G_k$	&$SU(4)$& $SU(N)$	&$U_A$	& $U_B$	& $U_R$	&$U_2$	& $U_3$	&$\ldots$	& $U_k$	\\ \hline
$J_1$&	\yF	&		&		&		& \yF		&		&$N-2$	& $+1/2$	& $1/2$	&	0	&	0	&		&	0	\\
$K_1$& \yA	&		&		&		&		&		& $-4$	&	$+1$	&	0	&	0	&	0	&		&	0	\\
$\P_2$& \yFb	&	\yF	&		&		&		&		&	0	&	$-1$	&	0	&	1	&	0	&		&	0	\\ \hline
$\P_3$& 		&	\yFb	&		&		&		&		&	0	&	$+1$	&	0	&	$-1$	&	1	&$\ldots$	&	0	\\
$\vdots$ &	&		& $\ddots$&	\yF	&		&		&	0	&$\vdots$	&	0	&$\vdots$&$\vdots$	&		&	0	\\ 
$\P_k$&		&		& 		&	\yFb	&		&	\yF	&	0	&$\pm1$	&	0	&$\mp1$	&$\pm1$	& $\ldots$	&	1	\\ \hline\hline
$U_1$	&	&		&		&		&		&		& $-2N$	& $-N/2$	& $0$	&	0	&	0	&		&	0	\\
$V_1$	&	&		&		&		&	\yA	&		& $0$	& $-N/2$	& $1$	&	0	&	0	&		&	0	\\
$\ww_1$	&	&		&		&		&		&		& $2N$	& $-N/2$	& $2$	&	0	&	0	&		&	0	\\ \hline
$X_1$	&	&		&		&		&	\yF	&		& $-N$	& $-N/2$	& $1/2$	&	0	&	0	&		&	0	\\
$Y_1$	&	&		&		&		&	\yFb	&		& $N$	& $-N/2$	& $3/2$	&	0	&	0	&		&	0	\\ \hline\hline
$Z_1$	&	&		&		&		&		&		& $0$	& $N$	& $0$	&	0	&	0	&		&	0	\\ \hline
\end{tabular}
\caption{Transformation properties of the composite fields in the confined phase of $G_1$, in the limit where $G_{2}\times \ldots \times G_k$ is weakly gauged.
The composite fields $U$, $V$, and $\ww$ exist only if $N$ is even; if $N$ is odd, then they are replaced by $X$ and $Y$.}  
\label{table:UVsymmetry2}
\end{table}

At the scale $\Lambda_2 < \Lambda_1$, the $G_2$ fields confine to form the following $G_1 \times G_2$ singlets:
\begin{align}
J_2 = (J_1 \P_2) &&
K_2 = (K_1 \P_2^2) &&
X_2 = (K_1^m J_1) &&
Y_2 = (K_1^{m-1} J_1^3)	\label{eq:IRops2}
\end{align}
\begin{align}
U_2 = (K_{1}^m) &&
V_2= (K_{1}^{m-1} J_{1}^2) &&
\ww_2 = (K_{1}^{m-2} J_{1}^4) &&
Z_2 = (\P_2^N).
\end{align}
The fields $J_2$ and $K_2$ transform under $G_3$ as $\yF$ and $\yA$ respectively. 

It is convenient to define the shorthand notation $B_i$, where $B_i=\{U_i, V_i, \ww_i\}$ for even $N=2m$, and $B_i=\{X_i, Y_i\}$ for odd $N=2m+1$. 
%The infrared operators are then $\{J_i, K_i, B_i, Z_i\}$ in both cases, as the definitions of $J$, $K$, and $Z$ do not depend on whether $N$ is even or odd.
At scales below $\Lambda_2$ and above $\Lambda_3$, the intermediate degrees of freedom are $\{ J_2, K_2, B_1, B_2, Z_1, Z_2, \P_3, \ldots, \P_k\}$.
This set of fields satisfies the anomaly matching conditions for $SU(4)_L \times SU(N)_R \times U(1)_A \times U(1)_B \times U(1)_R \times U(1)_3 \times \ldots \times U(1)_k$.

It is straightforward to continue this procedure until all groups including $G_k$ have confined, using the following recursive operator definition:
\begin{align}
J_{i} = (J_{i-1} \P_i) &&
K_i = (K_{i-1} \P_i^2) &&
X_i = (K_{i-1}^m J_{i-1}) &&
Y_i = (K_{i-1}^{m-1} J_{i-1}^3) 	\label{eq:IRopsi}
\end{align}
\begin{align}
U_i = (K_{i-1}^m) &&
V_i = (K_{i-1}^{m-1} J_{i-1}^2) &&
\ww_i = (K_{i-1}^{m-2} J_{i-1}^4) &&
Z_i = (\P_i^N).\label{eq:IRopsieven}
\end{align}
This definition can be applied to $i=1$ as well if we define $J_0 = Q$ and $K_0 = A$.
Below the scale $\Lambda_k$, all of the gauge groups have confined, and the approximate $U(1)_{i=1\ldots k}$ symmetries are broken to discrete $\ZZ_N$ groups. The charges under the remaining continuous family symmetries are shown in Table~\ref{table:UVsymmetryk}.

\begin{table}[h]
\centering
\begin{tabular}{| c | c c | c c c |} \hline
		&$SU(4)_L$	&$SU(N)_R$&$U_A$& $U_B$		& $U_R$		\\ \hline
$J_k$		& \yF		&	\yF	&$N-2$	& $\pm1/2$	& $1/2$		\\
$K_k$		&		&	\yA	& $-4$	&	$\pm1$	&	0		\\ \hline \hline
$U_\text{odd}$	&		&		& $-2N$	& $-N/2$		& $0$		\\
$V_\text{odd}$	&	\yA	&		& $0$	& $-N/2$		& $1$		\\
$\ww_\text{odd}$&		&		& $2N$	& $-N/2$		& $2$		\\ \hline
$U_\text{even}$&		&		& $-2N$	& $+N/2$		& $0$		\\
$V_\text{even}$&	\yA	&		& $0$	& $+N/2$		& $1$		\\
$\ww_\text{even}$&		&		& $2N$	& $+N/2$		& $2$		\\ \hline\hline
$X_\text{odd}$	&	\yF	&		& $-N$	& $-N/2$		& $1/2$		\\
$Y_\text{odd}$	&	\yFb	&		& $N$	& $-N/2$		& $3/2$		\\ \hline
$X_\text{even}$&	\yF	&		& $-N$	& $+N/2$		& $1/2$		\\
$Y_\text{even}$&	\yFb	&		& $N$	& $+N/2$		& $3/2$		\\ \hline\hline
$Z_\text{odd}$	&		&		& $0$	& $N$		& $0$		\\ 
$Z_\text{even}$	&		&		& $0$	& $-N$		& $0$		\\ \hline
\end{tabular}
\caption{The transformation properties of the composite fields in the fully confined phase of $SU(N)^k$ are shown. The subscript $B_\text{odd,even}$ refers to $i=1\ldots k$, whereas the baryon content $B_i = \{U_i, V_i, \ww_i\}$ or $B_i = \{X_i, Y_i\}$ depends on $N$.
The $U(1)_B$ charges of $J_k$ and $K_k$ are positive if $k$ is odd, and negative if $k$ is even. } 
\label{table:UVsymmetryk}
\end{table}

It must be shown that the basis of infrared operators is large enough to cover the moduli space. For the $SU(N)^k$ gauge group with fields $\{A, Q, \P_1, \ldots, \P_k\}$, the dimension of the moduli space is
\begin{equation}
\dim M_0(k) = \frac{N(N-1)}{2} + 4 N + k N^2 - k (N^2 - 1) = 4 N + \frac{N(N-1)}{2} + k , \label{eq:dimM0k}
\end{equation}
while the operator basis $\{J_k, K_k; B_1, \ldots , B_k; Z_1, \ldots, Z_k\}$ has dimension
\begin{equation}
N_\text{ops} = 4N + \frac{1}{2} N(N-1) + 9 k,
\end{equation}
 implying that there are $8 k $ complex constraints. 
By rearranging \eqref{eq:IRopsi} as follows, we can find $8(k-1)$ of the constraint equations:
\begin{align}
\begin{array}{rcl}
X_i &=& (K_{i-1}^m J_{i-1}) = (K_{i-2} \P_{i-1}^2)^m (J_{i-2} \P_{i-1}) = (K_{i-2}^m J_{i-2} ) (\P_{i-1}^{2m+1}) = X_{i-1} Z_{i-1}  \\
Y_i &=& (K_{i-1}^{m-1} J_{i-1}^3) = (K_{i-2} \P_{i-1}^2)^{m-1} (J_{i-2} \P_{i-1})^3 = (K_{i-2}^{m-1} J_{i-2}^3 ) (\P_{i-1}^{2m+1}) = Y_{i-1} Z_{i-1} , 
\end{array} \label{eq:classXYi}
\end{align}
for $i=2, 3 \ldots k$. Similarly,
\begin{align}
U_i = U_{i-1} Z_{i-1} &&
V_i = V_{i-1} Z_{i-1} &&
\ww_i = \uu_{i-1} Z_{i-1} . \label{eq:classUVWi}
\end{align}
The eight remaining constraints are provided by 
\begin{align}
X_k Z_k = K_k^{m} J_k  &&
Y_k Z_k = K_k^{m-1} J_k^3 , \label{eq:classXYk}
\end{align}
or
\begin{align}
U_k Z_k = \Pf (K_k) &&
V_k Z_k = K_k^{m-1} J_k^2 &&
\ww_k Z_k = K_k^{m-2} J_k^4 .
\end{align}
It is possible that these classical constraints may be quantum-modified.

\paragraph{Reduced operator basis:}

The classical constraints for $B_{i> 1}$ are mildly problematic, because Eqs.~(\ref{eq:classXYi}) and~(\ref{eq:classUVWi}) imply that these operators are redundant: that is, they can be written as products from a smaller operator basis, $\{B_1, Z_1, Z_2, \ldots Z_k\}$, and are therefore not independent degrees of freedom. Excitations of the $B_i$ fields above the vacuum acquire $\mathcal O(\Lambda_i)$ masses if they do not obey the classical constraints. These massive modes decouple at the scale $\Lambda_k$, leaving only the degrees of freedom consistent with the classical (or quantum-modified) constraints.
Unfortunately, anomaly cancelation depended on the fields $B_{i=2\ldots k}$: if these are not true degrees of freedom, then the anomaly matching conditions might not be satisfied. 

A solution to this problem can be seen by studying the $X_\text{odd}$ and $Y_\text{even}$ charges in Table~\ref{table:UVsymmetryk}. Their fermionic components have opposite charges under each of $U(1)_A$, $U(1)_B$, and $U(1)_R$. When we calculate the anomaly coefficients for each of the mixed and pure $U(1)$ anomalies, the contributions from each $X_\text{odd}$ cancel those from a $Y_\text{even}$ field. This is also true for the $SU(4)^2 U(1)$ and $SU(4)^3$ anomalies. Therefore, we refer to $X_\text{odd}$ and $Y_{\text{even}}$ as an ``anomaly neutral pair," indicating that they can be removed without changing any of the anomaly coefficients.
Similarly, $X_\text{even}$ and $Y_\text{odd}$ also form an anomaly neutral pair.

If $k$ is odd, then all of the operators $\{X_2, Y_2, \ldots, X_k, Y_k\}$ can be removed in neutral pairs. Substituting $X_k$ and $Y_k$ with their equations of motion, \eqref{eq:classXYk} becomes
\begin{align}
\left(X_1 Z_1 Z_2 \ldots Z_{k-1} \right) Z_k = K_k^{m} J_k  &&
\left(Y_1 Z_1 Z_2 \ldots Z_{k-1} \right) Z_k = K_k^{m-1} J_k^3 \label{eq:classXYred}
\end{align}
This is not possible if $k$ is even. To remove all the redundant operators, we must also remove a pair $\{X_1, Y_\text{even}\}$ or $\{X_\text{even}, Y_1\}$, and this is inconsistent: both $X_1$ and $Y_1$ are necessary to describe the moduli space. 

This can be seen if we move away from the origin along the flat direction parameterized by $(A^m Q)$, while keeping $\P_1=0$. Along this flat direction $X_1$ increases, but $X_\text{even} = 0$. Therefore, $X_1$ describes directions on the moduli space that cannot be described by $X_\text{even}$. Similarly, by increasing $(A^{m-1} Q^3)$ and fixing $\P_1=0$, we can see that $Y_1$ is just as necessary.

Quantum modification to \eqref{eq:classXYred} could explain why the odd $k$ and even $k$ situations are different. 
If $U(1)_B$ is broken in the vacuum, then $\{X_i, Y_i\}$ become an anomaly-neutral pair under the remaining symmetries, for any value of $i=1\ldots k$.
Based on $F=N$ \susy\ QCD, one would expect the classical relationships involving $\P_i$ and $\P_{i+1}$ to be quantum-modified. Specifically, the combination $(Z_{i-1} Z_i)$ has the same spurious $U(1)_i$ charge as $\Lambda_i^{b=2N}$, allowing modifications to equations such as \eqref{eq:classXYred}.
For example, the classical $k=4$ constraint for $X_4 Z_4$ might become
\begin{equation}
X_1 \left( Z_1 Z_2 Z_3 Z_4 + \beta_1 \Lb{2} Z_3 Z_4 + \beta_2 Z_1 \Lb{3} Z_4 + \beta_3 Z_1 Z_2 \Lb{4} + \beta_4 \Lb{2} \Lb{4} \right) = K_4^m J_4  \label{eq:qmXk4},
\end{equation}
with some as-yet-unknown coefficients $\beta_i$. As long as the coefficients are not zero, then the flat direction corresponding to $(A^m Q)\neq 0$ with $\P_1 = 0$ now requires some of the $Z_{i\neq 1}$ to have nonzero expectation values. 
In this $Z_1=0$, $X_1 \neq 0$ example, \eqref{eq:qmXk4} implies that $\Lb{2} (Z_3 Z_4 + \Lb{4} ) = 0$, spontaneously breaking $U(1)_B$ even in the limit where $\ev{X_1} \gg \Lambda_k$. Once $U(1)_B$ is broken in the vacuum, the operators $\{J_4, K_4, X_1, Y_1, Z_{i=1 \ldots 4} \}$ obey the anomaly matching conditions.

A quantum-modified constraint like \eqref{eq:qmXk4} also explains why $\{J_k, K_k, X_1, Y_1, Z_{i=1\ldots k} \}$ is consistent at the origin of moduli space if $k$ is odd. In this case the $Z_i = 0$ solution remains valid far away from the origin, because every $\Lb{}$ term multiplies at least one $Z$ field. Consider \eqref{eq:qmXk4} with $k=5$:
\begin{eqnarray}
K_5^m J_5 &=& 
X_1 \big( Z_1 Z_2 Z_3 Z_4 Z_5 +\beta_1 \Lb{2} Z_3 Z_4 Z_5 + \beta_2 Z_1 \Lb{3} Z_4 Z_5 +\beta_3 Z_1 Z_2 \Lb{4} Z_5 +\beta_4 Z_1 Z_2 Z_3 \Lb{5} \nonumber\\&&\ +\beta_5 Z_1 \Lb{3} \Lb{5} +\beta_6 \Lb{2} Z_3 \Lb{5} +\beta_7 \Lb{2} \Lb{4} Z_5 \big) \label{eq:qmXk5}.
\end{eqnarray}
In this case, the $(A^m Q) \neq 0$, $\P^N_{i=1\ldots k} =0$ flat direction remains on the moduli space for arbitrarily large values of $(A^m Q)$.

This does not mean that $U(1)_B$ is necessarily broken in the vacuum if $k$ is even. Let us fix $Z_i = 0$ for all $i=1\ldots k$ to ensure that $U(1)_B$ is not broken at the scale $\Lambda_i$. After imposing this constraint, \eqref{eq:qmXk4} becomes
\begin{equation}
X_1 = \frac{K_4^m J_4}{\Lb{2} \Lb{4} },
\end{equation}
implying that $X_1$ is not an IR degree of freedom when $U(1)_B$ is conserved. The same is true for $Y_1 \Lb{2} \Lb{4} = K^{m-1}_4 J^3_4$. In this particular vacuum $X_1$ and $Y_1$ are redundant operators, and after they are removed from the calculation the $U(1)_B$ anomaly coefficients match the ultraviolet theory.

Theories with even $N$ behave in essentially the same way.
Under the exact family symmetries, the operator pairs $\{U_\text{odd}, \ww_\text{even} \}$, $\{U_\text{even}, \ww_\text{odd} \}$, and $\{V_\text{odd}, V_\text{even} \}$ are anomaly-neutral.
%Recall that for $SU(4)$, $\yA = \yAb$.
As in the odd $N$ case, if $k$ is even then it is not possible to remove all the redundant $\{U_i,V_i,\ww_i\}$ operators while preserving the anomaly matching. This leads us to expect that the classical constraint equations
\begin{align}
U_{k} = U_1 \left( Z_1 Z_2 \ldots Z_{k-1} \right) &,&
V_{k} = V_1 \left( Z_1 Z_2 \ldots Z_{k-1} \right) &,&
\ww_{k} = \ww_1 \left( Z_1 Z_2 \ldots Z_{k-1} \right)
\end{align}
receive quantum modifications of the form
\begin{equation}
\Pf K_k = U_1 \left( Z_1 Z_2 \ldots Z_{k-1} + \ldots + (\Lb{2} \Lb{4} \ldots \Lb{k-2} ) Z_{k-1} Z_k + (\Lb{2} \Lb{4} \ldots \Lb{k}) \right) .
\end{equation}
if $k$ is even.
Either $U(1)_B$ is broken in the vacuum, or the operators $\{U_1, V_1, \ww_1\}$ are not degrees of freedom: in both cases, the IR theory satisfies t'~Hooft anomaly matching.
Thus, the reduced operator basis describes all infrared degrees of freedom, for both even and odd $N$.% The $B_{i \geq 2}$ operators can be integrated out below the scales $\Lambda_i$, as we will see from the dynamically generated superpotential. 

\subsection{Dynamically generated superpotential} \label{sec:orderedWd}

In this section we find a dynamically generated superpotential in the region of parameter space with $\Lambda_1 \gg \Lambda_2 \gg \ldots \gg \Lambda_k$. We begin by considering how the $W_d$ of \eqref{eq:w1odd} and \eqref{eq:w1even} becomes modified at the $G_2$ confinement scale. Ignoring the precise relative coefficients between terms,
\begin{eqnarray}
W^{(1)}_\text{odd} &=& \frac{1}{\Lb{1}} \left( X_1 Y_1 Z_1 - X_1 K_1^{m-1} J_1^3 - Y_1 K_1^{m} J_1 \right) \\
W^{(1)}_\text{even} &=& \frac{1}{\Lb{1}} \left( U_1 \ww_1 Z_1 - V_1^2 Z_1 - U_1 K_1^{m-2} J_1^4 + V_1 K_1^{m-1} J_1^2 -  \ww_1 K_1^m\right)
\end{eqnarray}
At the scale $\Lambda_2$, we expect $J_1$ and $K_1$ to confine to form the $B_2$ baryons. If we make these replacements in $W^{(1)}$, it becomes
\begin{eqnarray}
W^{(1)}_\text{odd} &=& \frac{1}{\Lb{1}} \left( X_1 Y_1 Z_1 - X_1 Y_2 - Y_1 X_2 \right) \\
W^{(1)}_\text{even} &=& \frac{1}{\Lb{1}} \left( U_1 \ww_1 Z_1 - V_1^2 Z_1-  \ww_1 U_2 - U_1 \ww_2 + V_1 V_2 \right)
\end{eqnarray}
It is likely that $G_1$ confinement changes the holomorphic scale $\Lambda_2$ to some new $\tilde{\Lambda}_2$. To find the relationship between $\Lambda_2$ and $\tilde{\Lambda}_2$, let us normalize the hadrons to have mass dimension $+1$:\footnote{Even after dividing by these powers of $\Lambda$, it is not necessarily true that the fields are canonically normalized. Corrections in the K\"{a}hler potential are likely to require additional normalization.}
\begin{align}
\tJ_1 = \frac{J_1}{\Lambda_1} &&
\tK_1 = \frac{K_1}{\Lambda_1^2} &&
%\tB_1 = \frac{B_1}{\Lambda_1^p} &&
\tZ_1 = \frac{Z_1}{\Lambda_1^{N-1}},
\end{align}
and similarly for the baryon operators $B_1$. The dynamically generated superpotential $W_2$ has the form
\begin{equation}
W^{(2)} = \sum_\text{contr.} \left(\frac{\tK_1^{N-2} \tJ_1^4 \P_2^N}{\tLb{2}}\right) = \sum_\text{contr.} \left(\frac{K_1^{N-2} J_1^4 \P_2^N}{\Lambda_1^{2N} \tLb{2}}\right).
\end{equation}
From \eqref{eq:wdgeneral}, symmetry requirements ensure that the superpotential has the form
\begin{equation}
W^{(2)} \sim \frac{A^{N-2} Q^4 \P_1^N}{\Lb{1}} \frac{\P_1^N \P_2^N}{\Lb{2}} \longrightarrow \frac{  K_1^{N-2} J_1^4 \P_2^N }{\Lb{1} \Lb{2}},
\end{equation}
allowing $\tLb{2}$ to be expressed as
\begin{equation}
\tilde{\Lambda}_2^{2N-1} =\frac{1}{\Lambda_1}  \Lambda_2^{2N} .
\end{equation}
This expression can also be derived with the same result by matching the gauge couplings at the mass threshold $\Lambda_1$. Based on this agreement, we do not expect the superpotential $W_2$ to receive modifications of the form
\begin{equation}
W^{(2)} \rightarrow \left(1+ \frac{Z_1 Z_2}{\Lb{2}} + \ldots \right) W^{(2)},
\end{equation}
even though such terms are consistent with the family symmetries.

As confinement continues, the products of intermediate mesons $J_2$ and $K_2$ can be replaced with $G_3$ baryons. Each $i=1\ldots k$ superpotential $W^{(i)}$ becomes
\begin{eqnarray}
W^{(i<k)}_\text{odd} &=& \left(\prod_{j=1}^{i} \Lb{j} \right)^{-1} \left( X_i Y_i Z_i - X_i Y_{i+1} - Y_i X_{i+1} \right) \\
 W^{(k)}_\text{odd} &=& \left(\prod_{j=1}^k \Lb{j} \right)^{-1} \left( X_k Y_k Z_k - X_k K_k^{m-1} J_k^3 - Y_k K_k^{m} J_k \right), \\
W^{(i<k)}_\text{even} &=& \left(\prod_{j=1}^{i} \Lb{j} \right)^{-1} \left( U_i \ww_i Z_i - V_i^2 Z_i-  \ww_i U_{i+1} - U_i \ww_{i+1}+ V_i V_{i+1} \right)  \\
W^{(k)}_\text{even} &=& \left(\prod_{j=1}^k \Lb{j} \right)^{-1} \left( U_k \ww_k Z_k - V_k^2 Z_k-  \ww_k K_k^m - U_k K_k^{m-2} J_k^4 + V_k K_k^{m-1} J_k^2 \right).
\end{eqnarray}
The full superpotential is the sum
\begin{equation}
W_d =  \sum_{i=1}^{k} W^{(i)} . \label{eq:fullWd}
\end{equation}
%up to quantum corrections of the form
%%Taken together, the equations of motion from $W_d$ enforce the $8k$ classical constraints listed in Section~\ref{sec:infraops}. 
%%This superpotential is not the most general superpotential allowed by the symmetries. In particular, the combination $(Z_{i} Z_{i+1}/ \Lb{i+1})$ is neutral under all of the family symmetries, including the approximate anomalous ones. For example, there is nothing preventing the $W^{(i)}$ from receiving corrections of the form
%\begin{equation}
%W^{(i)} \rightarrow W^{(i)}+ \beta  W^{(i)} \frac{Z_1 Z_2}{\Lb{2}}  + \ldots .
%\end{equation}
%%unless $\beta$ is such that the equations of motion no longer approach the classical results in the weak coupling $\Lambda_i \rightarrow 0$ limit.
%\rev{We ignore this possibility for now,} and consider the equations of motion that follow from \eqref{eq:fullWd}.

\paragraph{Equations of motion:}

%In the $k=1$ case we were able to adjust the phase of $W_d$ by performing a $U(1)_R$ rotation. Although this is still a symmetry of $W_d$ for $k>1$, it is not possible to remove the relative phases between the various $W^{(i)}$. These are controlled by the $\theta_i$ parameters, and may have physical significance.

%From Equations~\ref{eq:w1odd} and \ref{eq:w1even}, recall that there is an overall undetermined factor $\alpha$ in each of the $W^{(i)}$ superpotentials. This factor did not affect the equations of motion for the $k=1$ case: however, if the factors $\alpha_i$ in each $W^{(i)}$ are different, then they will appear in the equations of motion. We may avoid this problem by absorbing these unknown factors into the definition of the holomorphic scales $\Lambda_i$, as long as $W_d$ has the form \eqref{eq:fullWd}. 

Let us consider equations of motion of the form $\partial W / \partial B_1$, where $B_1 = \{U_1, V_1, \ww_1, X_1, Y_1\}$ is any of the $G_1$ baryons.
It is easy to show that these equations are
\begin{align}
Y_2 = Y_1 Z_1 &&
X_2 = X_1 Z_1 &&
X_1 Y_1 = 0
\end{align}
for odd $N$, and
\begin{align}
\ww_2 = \ww_1 Z_1 &&
V_2 = V_1 Z_1 &&
U_2 = U_1 Z_1 &&
U_1 \ww_1 = V_1^2
\end{align}
for even $N$. The $\partial W/\partial B_2$ equations yield more surprising results: for example, %because $X_2$ appears in both $W^{(1)}$ and $W^{(2)}$, its equations of motion produce
%\begin{eqnarray}
%\frac{\partial W_d}{\partial X_2 } = -\frac{Y_1}{\Lb{1}}  + \frac{Y_2 Z_2}{\Lb{1} \Lb{2}} &=& 0  \\
%Y_2 Z_2 - Y_3  &=&  Y_1 \Lb{2} .
%\end{eqnarray}
\begin{eqnarray}
\frac{\partial W_d}{\partial X_2 } = -\frac{Y_1}{\Lb{1}}  + \frac{Y_2 Z_2}{\Lb{1} \Lb{2}} =0   &\longrightarrow&
Y_2 Z_2 - Y_3  =  Y_1 \Lb{2} .
\end{eqnarray}
The classical constraint $Y_2 Z_2 = Y_3$ is modified, due to the appearance of $X_2$ in both $W^{(1)}$ and $W^{(2)}$. For $i=2, 3 \ldots (k-1)$, we find
\begin{equation}
B_i Z_i = B_{i+1} + \Lb{i} B_{i-1} .
\end{equation}
The equations of motion $\partial W_d /\partial Z_i$ are not modified, so that
\begin{align}
X_i Y_i = 0 &,& U_i \ww_i = V_i^2 
\end{align}
for all $i$.
Finally, the $B_k$ equations of motion are
\begin{align}
X_k Z_k = K_k^{m} J_k + \Lb{k} X_{k-1} &,&
Y_k Z_k = K_k^{m-1} J_k^3 + \Lb{k} Y_{k-1} 
\end{align}
for odd $N$, and
%\begin{align}
%\begin{array}{rcl}
%U_k Z_k &=& K_k^m + \Lb{k} U_{k-1} \\
%V_k Z_k &=& K_k^{m-1} J_k^2 + \Lb{k} V_{k-1} \end{array} &&
%\begin{array}{rcl}
%\ww_k Z_k &=& K_k^{m-2} J_k^4 + \Lb{k} \ww_{k-1} \\
%U_k \ww_k &=& V_k^2 \end{array}
%\end{align}
\begin{align}
U_k Z_k = K_k^m + \Lb{k} U_{k-1} &,&
V_k Z_k = K_k^{m-1} J_k^2 + \Lb{k} V_{k-1} &,&
\ww_k Z_k = K_k^{m-2} J_k^4 + \Lb{k} \ww_{k-1}
\end{align}
for even $N$.

%In the product group case, the phases of $\Lambda^b_i$ now affect the equations of motion. 
Recall from Section~\ref{sec:antiprops} that each gauge group $SU(N)_i$ has a related $CP$ parameter $\theta_i$, which determines the phase of the holomorphic scale $\Lb{i}$. %In the UV theory with $W=0$, the $U(1)_i$ symmetry is approximately conserved: however, at $\mathcal O(\Lambda_i)$ scales, shifts in $\theta_i$ become important and $U(1)_i$ is broken.
%The dynamically generated superpotential $W_d^{(i)}$ breaks the approximate $U(1)_i$ symmetry to a discrete $\ZZ_N$ subgroup.
%For each $U(1)_i$ there is a conserved $\ZZ_N$ subgroup in the infrared: although the field $\P_i$ has charge $+1$ under $U(1)_i$, the superpotential is proportional to $\P_i^N$, and is therefore invariant under the transformation $\theta_i \rightarrow \theta_i + \frac{2\pi}{N}$. The transformations of the hadrons under the $(\ZZ_N)^k$ family symmetry can be inferred from Table~\ref{table:UVsymmetry}.
Although $\Lambda^b$ did not appear in the $k=1$ equations of motion, the phases of $\Lambda^b_i$ do affect the equations of motion in the product group case. The overall phase of $W_d$ can still be removed by performing a $U(1)_R$ rotation; however, the relative phases between the $\Lambda_i$ may have physical effects.

Armed with these iterative equations of motion, we can rewrite the larger baryons $B_{i>1}$ in terms of $\{B_1\} $ and the $Z_i$ fields.
For example,
\begin{eqnarray}
B_2 &=& B_1 Z_1 \\
B_3 &=& B_1 (Z_1 Z_2 - \Lb{2} )  \label{eq:qmB3} \\
B_4 &=& B_1 (Z_1 Z_2 Z_3 - \Lb{2} Z_3 - Z_1 \Lb{3} ) \label{eq:qmB4} \\
B_5 &=& B_1 (Z_1 Z_2 Z_3 Z_4 - \Lb{2} Z_3 Z_4 - Z_1 \Lb{3} Z_4 - Z_1 Z_2 \Lb{4} + \Lb{2} \Lb{4} ).
\end{eqnarray}
Our guesses in Eqs.~\ref{eq:qmXk4} and \ref{eq:qmXk5} as to the form of the quantum modification are correct, with $\beta_i = \pm 1$ for each coefficient.
%As we do not make assumptions about the values of $\theta_i$, the phase of $\Lb{i}$ remains undetermined. %: if we wish, we can absorb these factors of $(-1)$ into our definition of $\theta_i$.
This process is extended to arbitrary $B_i$ in the following way: each classical constraint involving products of the form $(Z_1 Z_2 \ldots Z_j)$ is modified by replacing adjacent pairs $(Z_{i-1} Z_i)$ by $(-\Lb{i})$, and each possible term is added to the product $(Z_1 \ldots Z_j)$. 
%It is convenient to introduce a shorthand notation %%but we won't until the next paper.
After making these adjustments, the $k^\text{th}$ equations of motion return the following constraints if $k$ is odd:
\begin{align}
\begin{array}{rcl}
K_k^m J_k &=& X_1 \big\{ ( Z_1  \ldots Z_k) - \Lb{2} (Z_3 \ldots Z_k) +  \ldots +(-1)^{(k-1)/2} (\Lb{2} \Lb{4} \ldots \Lb{k-1}) Z_k \big\} \\
K_k^{m-1} J_k^3 &=& Y_1 \big\{ ( Z_1  \ldots Z_k) - \Lb{2} (Z_3 \ldots Z_k) +  \ldots + (-1)^{(k-1)/2} (\Lb{2} \Lb{4} \ldots \Lb{k-1}) Z_k \big\}, \\% \end{array}\\ \begin{array}{rcl}
K_k^m &=& U_1 \big\{ ( Z_1  \ldots Z_k) - \Lb{2} (Z_3 \ldots Z_k) +  \ldots + (-1)^{(k-1)/2} (\Lb{2} \Lb{4} \ldots \Lb{k-1}) Z_k \big\}   \\
K_k^{m-1} J_k^2 &=& V_1 \big\{ ( Z_1  \ldots Z_k) - \Lb{2} (Z_3 \ldots Z_k) +  \ldots + (-1)^{(k-1)/2} (\Lb{2} \Lb{4} \ldots \Lb{k-1}) Z_k \big\}  \\
K_k^{m-2} J_k^4 &=& \ww_1 \big\{ ( Z_1  \ldots Z_k) - \Lb{2} (Z_3 \ldots Z_k) +  \ldots + (-1)^{(k-1)/2} (\Lb{2} \Lb{4} \ldots \Lb{k-1}) Z_k \big\}   ,
\end{array} \label{eq:finalqmkodd}
\end{align}
or if $k$ is even:
\begin{align}
\begin{array}{rcl}
K_k^m J_k &=& X_1 \big\{ ( Z_1  \ldots Z_k)  +  \ldots -(-1)^{\frac{k}{2}} (\Lb{2} \ldots \Lb{k-2} ) Z_{k-1} Z_k + (-1)^{\frac{k}{2}} (\Lb{2} \Lb{4} \ldots \Lb{k})  \big\}  \\
K_k^{m-1} J_k^3 &=& Y_1 \big\{ ( Z_1  \ldots Z_k)  +  \ldots -(-1)^{\frac{k}{2}} (\Lb{2} \ldots \Lb{k-2} ) Z_{k-1} Z_k + (-1)^{\frac{k}{2}} (\Lb{2} \Lb{4} \ldots \Lb{k})  \big\} ,\\% \end{array}\\ \begin{array}{rcl}
K_k^m &=& U_1 \big\{ ( Z_1  \ldots Z_k)  +  \ldots -(-1)^{\frac{k}{2}}(\Lb{2} \ldots \Lb{k-2} ) Z_{k-1} Z_k + (-1)^{\frac{k}{2}} (\Lb{2} \Lb{4} \ldots \Lb{k})  \big\} \\
K_k^{m-1} J_k^2 &=& V_1 \big\{ ( Z_1  \ldots Z_k)  +  \ldots -(-1)^{\frac{k}{2}} (\Lb{2} \ldots \Lb{k-2} ) Z_{k-1} Z_k + (-1)^{\frac{k}{2}} (\Lb{2} \Lb{4} \ldots \Lb{k})  \big\}  \\
K_k^{m-2} J_k^4 &=& \ww_1 \big\{ ( Z_1  \ldots Z_k)  +  \ldots -(-1)^{\frac{k}{2}}(\Lb{2} \ldots \Lb{k-2} ) Z_{k-1} Z_k +(-1)^{\frac{k}{2}} (\Lb{2} \Lb{4} \ldots \Lb{k})  \big\} .
\end{array} \label{eq:finalqmkeven}
\end{align}
In both cases, the origin of moduli space is a solution to the equations of motion. 

As we suggested in Section~\ref{sec:infraops}, if $k$ is even then the $B_1$ fields are not independent degrees of freedom when $Z_{i=1\ldots k} = 0$: 
\begin{align}
%\begin{array}{rcl}
%K_k^m J_k &=& X_1 (-1)^{\frac{k}{2}} (\Lb{2} \Lb{4} \ldots \Lb{k})   \\
%K_k^{m-1} J_k^3 &=& Y_1  (-1)^{\frac{k}{2}} (\Lb{2} \Lb{4} \ldots \Lb{k})  \end{array} &;&
\begin{array}{rcl}
K_k^m &=& U_1  (-1)^{\frac{k}{2}} (\Lb{2} \Lb{4} \ldots \Lb{k}) \\
K_k^{m-1} J_k^2 &=& V_1 (-1)^{\frac{k}{2}} (\Lb{2} \Lb{4} \ldots \Lb{k})   \\
K_k^{m-2} J_k^4 &=& \ww_1 (-1)^{\frac{k}{2}} (\Lb{2} \Lb{4} \ldots \Lb{k})  
\end{array} 
&;&
\begin{array}{rcl}
K_k^m J_k &=& X_1 (-1)^{\frac{k}{2}} (\Lb{2} \Lb{4} \ldots \Lb{k})   \\
K_k^{m-1} J_k^3 &=& Y_1  (-1)^{\frac{k}{2}} (\Lb{2} \Lb{4} \ldots \Lb{k})  .\end{array} 
\label{eq:qmkevenZ0}
\end{align}
Therefore, if $U(1)_B$ is a symmetry of the vacuum and $k$ is even, then the $B_1$ fields are completely determined by $J_k$ and $K_k$. After removing the $B_1$ fields, the t'~Hooft anomaly matching conditions are satisfied.
Elsewhere on the moduli space the $B_1$ fields may vary independently from $K_k$ and $J_k$, $U(1)_B$ is spontaneously broken by $\ev{Z_i} \neq 0$, and the anomaly coefficients for the infrared symmetries match the values calculated in the ultraviolet theory.

%In the odd $k$ case, the $B_1$ fields are degrees of freedom even in the neighborhood of the origin: in particular, along the $Z_{i=1\ldots k} =0$ directions every $B_1$ can be varied independently from $J$ and $K$. All of the fields $\{J_k, K_k, B_1, Z_1 \ldots Z_k\}$ are necessary for the anomaly matching.

%%%%MOVE TO END OF SECTION
%
%We now have a consistent description of the $SU(N)^k$ theory everywhere on the moduli space, for odd or even $k$, as long as the $G_1 \times \ldots \times G_k$ confinement proceeds in order from $i=1$ to $i=k$.
%The sets of gauge invariant infrared degrees of freedom obey anomaly matching conditions; the origin remains on the quantum moduli space; and there is a dynamically generated superpotential. The quantum and classical moduli spaces do not match: for even $k$, the differences are particularly significant in the vicinity of the origin, which is not true for odd $k$. 
%Even so, it is possible for the theory to confine without chiral symmetry breaking for any value of $k$, and all of the interactions are determined by $W_d$. Therefore, in the ordered parameter limit, we say that the $SU(N)^k$ extension to $A+4Q + N \barQ$ is s-confining.

\subsection{Additional tests} \label{sec:disorderedWd}

So far we have restricted our attention to the ordered $\Lambda_1 > \ldots > \Lambda_k$ case to find the dynamically generated superpotential. Due to the holomorphy of the superpotential, changes in the $\Lambda_i$ hierarchy should not alter the form of the superpotential. In this section we test this supposition by considering the $\Lambda_1 \ll \Lambda_{i\neq 1}$ case. In this limit the $SU(N)^k$ model reduces to an $SU(N)^{k-1}$ extension to $F=N$ \susy\ QCD which has been studied by Chang and Georgi~\cite{Chang:2002qt}. 

As $\Lambda_1 \rightarrow 0$, the $A$ and $Q$ fields decouple from the strongly coupled $\P_i$.
Chang and Georgi find that the infrared operators involving only $\P_i$ obey the following constraints:
%have the same form as our equations of motion for $B_{i\geq 2}$: for example, 
\begin{eqnarray}
\det (\P_1 \P_2) &=& Z_1 Z_2 - \Lb{2}  \\
\det (\P_1 \P_2 \P_3) &=& Z_1 Z_2 Z_3 - \Lb{2} Z_3 - Z_1 \Lb{3}  \\
\det (\P_1 \P_2 \P_3 \P_4) &=& Z_1 Z_2 Z_3 Z_4 - \Lb{2} Z_3 Z_4 - Z_1 \Lb{3} Z_4 - Z_1 Z_2 \Lb{4} + \Lb{2} \Lb{4} ,
\end{eqnarray}
and so on. This is exactly the same form we derived for $B_{i\geq 2}$ in Section~\ref{sec:orderedWd}. 
At scales above $\mathcal O(\Lambda_1)$ but below $\Lambda_{i>1}$, the $G_1$ charged degrees of freedom include $A$, $Q$, and $M = (\P_1 \P_2 \ldots \P_k)$. Let us define the mass-normalized field $\tM$,
\begin{equation}
\tM = \frac{(\P_1 \P_2 \ldots \P_k)}{\Lambda_2 \Lambda_3 \ldots \Lambda_k},
\end{equation}
and let the fields $\{A, Q, \tM\}$ confine under $G_1$, producing
\begin{align}
J_k = Q \tM &,& K_k = A \tM^2 &,& Z_M = \det (\tM) ,
\end{align} 
and the baryons $B_1= \{U_1, V_1, \ww_1; X_1, Y_1\}$ as defined in Section~\ref{sec:infraops}.
The dynamically generated superpotential is
\begin{eqnarray}
W_\text{odd} &=& \frac{X_1 Y_1 Z_M - X_1 K_k^{m-1} J_k^3 - Y_1 K_k^m J_k }{\tLb{1}} \\
W_\text{even} &=& \frac{(U_1 \ww_1 - V_1^2) Z_M - U_1 K_k^{m-2} J_k^4 + V_1 K_k^{m-1} J_k^2 - \ww_1 K_k^m }{\tLb{1}} .
\end{eqnarray}
The effective scale $\tLb{1}$ contains a product of $(\P_1^N \ldots \P_k^N)$ and $\Lb{2} \ldots \Lb{k}$, so that the superpotential is invariant under the spurious symmetries.

There is also a quantum modified constraint
\begin{equation}
Z_M = \det \tM = (Z_1 \ldots Z_k) - \Lb{2} (Z_3 \ldots Z_k) + \{ \text{all other contractions} \}. \label{eq:zmqd}
\end{equation}
If we use a Lagrange multiplier $\lambda$, \eqref{eq:zmqd} follows from the superpotential
\begin{equation}
W'_d = \lambda \left\{ Z_M -  (Z_1 \ldots Z_k) +  ( \text{all contractions} ) \right\} .
\end{equation}
After replacing $Z_M$ with $\{Z_i\}$, the equations of motion are identical to Eqs.~(\ref{eq:finalqmkodd}) and~(\ref{eq:finalqmkeven}), suggesting that there is no phase transition in the parameter space. 

Notice that the equations of motion from $Z_M$ also determine a vacuum solution for $\lambda$:
\begin{eqnarray}
\frac{\partial W_\text{odd}}{\partial Z_M} &=& \frac{X_1 Y_1}{\tLb{1}} + \lambda = 0 \\
\frac{\partial W_\text{even}}{\partial Z_M} &=& \frac{U_1 \ww_1 - V_1^2}{\tLb{1}} + \lambda = 0 \\
\end{eqnarray}
Thus, the Lagrange multiplier can be treated as a new redundant baryon operator, which should be integrated out along with the other redundant fields.
%Now that $\Lambda_1 \ll \Lambda_{i\neq 1}$, the redundant operators $B_{i>1}$ are not generated: instead, depending on the confinement order within $\{ \Lambda_{2\ldots k} \}$, we generate operators of the form $Z_{ij} \equiv \det(\P_i \P_{i+1} \ldots \P_j)$. Each of these redundant operators can be replaced with their equations of motion, as derived in~\cite{Chang:2002qt}.

Finally, let us consider regions of parameter space in which $\Lambda_1$ is neither the largest nor the smallest confinement scale. In these cases the redundant operators include a mix of $B_i$ and $Z_{ij}$, all of which produce the same equations of motion in the reduced operator basis.
%In every permutation of the confinement order, the set of naturally generated ``non-minimal" operators ($B_{i>1}$, $Z_{i, i+1 \ldots}$, and $\lambda_i$) changes, but the minimal basis $\{J_k, K_k, B_1, Z_1 \ldots Z_k\}$ remains the same.
For any arrangement, at the last confinement scale $\Lambda_f$ there is a dynamically generated superpotential of the form
\begin{equation}
W^{(f)} \sim \frac{\tK_f^{N-2} \tJ_f^4 \tM^N }{\tLb{f} }, 
\end{equation}
%~~~\text{where } ~~
%\end{align}
%where
%\begin{align}
%J_k = (J_f M) , ~~~
%K_k = (K_f M^2) ,
%\end{align}
where $J_f$, $K_f$, and $M$ are such that 
\begin{align}
(J_f M) = (Q \P_1 \ldots \P_f)(\P_{f+1} \ldots \P_k) = J_k  ,&&
(K_f M^2) = \left(A \P_1^2 \ldots \P_f^2\right)\! \! \left(\P_{f+1} \ldots \P_k^2\right)^2 = K_k,
\end{align}
and where $\{ \tJ_f, \tK_f, \tM\}$ are normalized to have mass dimension $+1$.
Under the remaining gauged $G_f$, these fields satisfy the index condition for s-confinement, $\sum_j \mu_j - \mu_G = 2$, and there is a dynamically generated superpotential.
Lagrange multipliers $\lambda_i$ enforce the constraint between the operators $\det (\P_i \ldots \P_j)$ and $\{Z_i \ldots Z_j \}$, and the equations of motion provide a relationship between $\lambda_i$ and the other hadrons. 
%There is a subtlety here: the equations of motion depend on the effective scales $\tLb{i}$, which in turn contain factors of $\ev{Z_j}$ and $\Lb{j}$. One must be careful when matching the holomorphic scales $\tLb{i}$ to the original, high-energy $\Lb{i}$.
After replacing the redundant operators with their equations of motion, we find that the constraints relating $\{J_k, K_k\}$ to $\{B_1, Z_i\}$ are unchanged.

\paragraph{Flow:}

It is a necessary condition for s-confining theories that their description in terms of gauge-invariants is valid in the Higgs phase, when some fields acquire large expectation values and spontaneously break the gauge group to a subgroup. If the low-energy theory does not s-confine, then the original theory cannot be s-confining either. This is the ``flow requirement" of~\cite{Csaki:1996sm}, which we use in this section to test the $SU(N)^k$ theory.%In this section we describe our results: a more detailed treatment is found in Appendix~\ref{sec:flow}.

%There are two directions in which the $SU(N)^k$ group can be broken.

In the $\ev{J_k}^i_j \gg \Lambda$ vacuum with $\ev{A_{\alpha\beta}} = 0$, the $SU(N)^k$ group is broken to $SU(N-1)^k$ in the classical limit. 
%An $SU(4)_L \times SU(N)_R$ rotation can rotate the large VEV into the $(i=4,j=N)$ component.
This requires a nonzero $(\P_i)^\alpha_\beta$ for every $\P_i$, which break each gauged $SU(N)_i$ to $SU(N-1)_i$. The $SU(N)_i\times SU(N)_{i+1}$ bifundamentals $\P_i$ decompose into $SU(N-1)\times SU(N-1)$ representations as follows:
\begin{align}
SU(N) \times SU(N) \rightarrow SU(N-1) \times SU(N-1): &&
{(\yFb, \yF) }\longrightarrow (\yFb, \yF) \oplus (\yFb, \yI) \oplus (\yI, \yF) \oplus (\yI,\yI).
\end{align}
The $(2N-1)$ broken generators of each gauge group $G_{i\neq1}$ ``eat" the combination $\yF+\yFb+\yI$ from $\P_{i-1}$ and $\P_i$ to create $(2N-1)$ massive gauge superfields, leaving behind the $(\yFb, \yF)$ bifundamental fields.

The $G_1$ group behaves somewhat differently: its broken generators ``eat" the $(\yFb, \yI)$ part of $\P_1$ and a linear combination of the $\yF$ superfields $Q_{i=1\ldots 4}$. % leaving $3 Q' + (N-1) \P_1'$ charged under $SU(N-1)_1$. 
Under $SU(N-1)_1$ the $\yA$ field decomposes as $(\yA \oplus \yF)$, so that the ``eaten" $Q$ field is replaced by a component of $A$. 
After removing the massive superfields, the $SU(N-1)_1$ charged matter is $A' + 4Q' + (N-1) \P_1'$. %, with a sequence of $SU(N-1)_{i} \times SU(N-1)_{i+1}$ bifundamentals $\P_i'$. so that the overall effect on the $SU(N)^k$ model is to replace $N$ with $N-1$.
The overall effect of $\ev{J_k}\gg \Lambda$ on the $SU(N)^k$ model is to replace $N$ with $N-1$.
%Eventually, the gauge group can be broken in this way to $SU(3)^k$.  

Now let us consider the limit where $\ev{A_{\alpha\beta}}\gg\Lambda$ and $\ev{J}=0$. In the even $N=2m$ case with $\ev{U_1 = \Pf A} \gg \Lambda_1$, $SU(2m)_1$ is broken to $Sp(2m)_1$
and $\yA$ decomposes into $\yA_{Sp} \oplus \yI$. Here $\yA_{Sp}$ is the $(2m^2 - m -1)$ dimensional representation of $Sp(2m)$. There are also $(2m^2 - m -1)$ broken $SU(2m)$ generators, so the superfield $A' = \yA_{Sp}$ is eaten. 

The fields $Q$ and $\P_i$ are not directly affected by $\ev{\Pf A}$: however, as $Sp(2m)$ has no complex representations, $Q$ and $\P_1$ are effectively $(2m+4)$ quarks charged in the $\yF$ representation of $Sp(2m)$. This theory is known to s-confine~\cite{Intriligator:1995ne}. It is likely that the $Sp(2m) \times SU(2m)^{k-1}$ product group theory is also s-confining: we explore this possibility in Section~\ref{sec:spinextension}.

In the case where $N$ is odd, %there is no gauge invariant $\Pf A$ operator. Instead, 
an expectation value $\ev{X_1} = \ev{A^m Q} \gg \Lambda$ breaks $SU(2m+1)$ to $Sp(2m)$ instead. Aside from a few extra singlets and massive gauge bosons, there is little difference between the odd $N$ and even $N$ cases: the infrared theory is $Sp(2m) \times SU(2m)^{k-1}$.

\paragraph{Conclusion:}

Our product group extension to the $A+4Q + N \barQ$ model exhibits the behavior required for an s-confining theory. 
The set of gauge invariant operators $\{J_k, K_k, B_1, Z_{1\ldots k}\}$ satisfies the t'~Hooft anomaly matching conditions; the origin remains on the quantum moduli space, so the theory can confine without breaking chiral symmetry; and there is a dynamically generated superpotential. Furthermore, the operators $\{J_k, K_k, B_1, Z_{1\ldots k}\}$ provide a smooth description of the entire moduli space: there is no gauge invariant order parameter to distinguish the confined and Higgs phases. By considering the flow along flat directions, we have also found another product group extension to an s-confining theory, $Sp(2m) \times SU(N)^{k-1}$.

%We now have a consistent description of the $SU(N)^k$ theory everywhere on the moduli space, for odd or even $k$. 
%%, as long as the $G_1 \times \ldots \times G_k$ confinement proceeds in order from $i=1$ to $i=k$.
%The gauge invariant infrared degrees of freedom obey the anomaly matching conditions; the origin remains on the quantum moduli space; and there is a dynamically generated superpotential. 
%The quantum and classical moduli spaces do not match for $k>1$, but it remains possible for the theory to confine without chiral symmetry breaking. Therefore, in the ordered parameter limit, we say that the $SU(N)^k$ extension to $A+4Q + N \barQ$ is s-confining.

%%%%%DISCUSS? Symmetry preserving Higgs phase 

%%%%%%%%%%%%%%%%%%%

%\subsection{Special Case: $SU(4)$}

\section{Other S-Confining Theories} \label{sec:misc}

In the previous section we find strong evidence that the product group extension to the $A+4Q +N\barQ$ model is s-confining. In this section we consider the follow-up question: how many other s-confining models can be extended into product groups? We have already suggested that $Sp(2m)$ with $(2m+4) \yF$ can be extended into an $Sp(2m)\times SU(N)^{k-1}$ product group model. If this theory is not s-confining, then the $SU(N)^k$ $A+4Q + N \barQ$ model is not s-confining either. We discuss the behavior of this theory in Section~\ref{sec:spinextension}.

There are also additional possibilities for the $A+4Q + N \barQ$ model in the case where $N=4$. In this special case the entire $SU(4)_L\times SU(N)_R$ family symmetry can be gauged: we consider whether or not such theories are s-confining in Section~\ref{sec:su4LR}.
In Sections~\ref{sec:sqcdp} and~\ref{sec:notsconf} we discuss the other s-confining theories in~\cite{Csaki:1996zb} with family symmetries large enough to accommodate a gauged $SU(N)$ subgroup. This includes \susy\ QCD with $F=N+1$ flavors, and $Sp(2m)$ with $(\yA + 6 \yF)$ matter for $m=2$ and $m=3$. We show that some of these theories are not s-confining.
%This is true for $A+4Q + N\barQ$, because the $G_2$ index sum $(\sum_j \mu_j - \mu_G)$ increases from $0$ to $+2$ when $G_1$ confines, thus allowing the hadrons to s-confine under $G_2$.
 
Due to the lack of an index constraint on the matter content, it is difficult to conduct a systematic search for new s-confining product groups. We have seen in the $A+4Q + N\barQ$ model that $G_1$ confinement increases the index sum of the $G_2$ charged matter by $+2$, but other confining theories tend to change the index sum by varying amounts. Therefore, the list of theories considered in this section is presumably incomplete.
%It is plausible that other s-confining theories exist with entirely different forms: for example, it may be possible to replace
%%Other confining theories tend to change the index sum by different amounts, making it difficult to conduct a systematic search for other s-confining product groups.
%It may also be possible to extend any of the known s-confining theories using more complicated matter content, or more complicated product groups.

We restrict our attention to s-confining models which can be extended by gauging a subgroup of the family symmetries and adding bifundamental fields.
%In Section~\ref{sec:su4LR} we consider the $N=4$ special case for the $A+4Q + N\barQ$ model, in which both $SU(4)_L$ and $SU(4)_R$ are gauged.
%In the sections that follow, we consider the $Sp(2m)$ and \susy\ QCD models. 
Our goal is to determine whether product group s-confinement is possible in each model, based on the index constraint after confinement. This is sufficient to show which of the product group extensions are obviously not s-confining. A more detailed analysis is appropriate for the theories which pass this test.

\subsection{Special case: $SU(4)$} \label{sec:su4LR}

In this section, we extend the $N=4$ $A+4Q + N\barQ$ model by gauging $SU(4)_L^\ell \times G_0 \times SU(4)_R^\arr$ for some $\ell$ and $\arr$. Here $G_0$ is the $SU(4)$ gauge group containing the $\yA + 4(\yF + \yFb)$ matter, and every other gauged $SU(4)$ contains four flavors of $(\yF + \yFb)$. 
It is convenient to relabel the hadrons to reflect the $Q \leftrightarrow \P$ symmetry of the matter content of the $A+4 Q + 4 \barQ$ model:
\begin{align}
M = Q \barQ ,&&
\barK = A \barQ^2 ,&&
K = A Q^2 ,&&
U = A^2 ,&&
Z = Q^4 ,&&
\barZ = \barQ^4 .
\end{align}
A convenient redefinition of the $U(1)_A \times U(1)_B \times U(1)_R$ charges is shown in Table~\ref{table:UVmatterl2r2}, for $\ell=\arr=2$.

\begin{table}[h]
\centering
\begin{tabular}{| c | c | c c c c c | c | c c c |} \hline
   	& $SU(4)_L$	& $G_2$ 	& $G_1$	& $G_0$	&$\bG_1$	&$\bG_2$	&$SU(4)_R$	& $U_A$	&$U_B$	&$U_R$	\\ \hline
$Q_2$	& \yFb	&	\yF	&		&		&		&		&			&	$1$	&	$0$	&	0	\\
$Q_1$	&		& \yFb	&	\yF	&		&		&		&			&	$-1$	&	0	&	0	\\
$Q_0$	&		& 		&	\yFb	&	\yF	&		&		&			&	$1$	&	$0$	&	0	\\ 
$A$		&		& 		&		&	\yA	&		&		&			&	$-2$	&	$-2$	&	1	\\
$\P_0$	&		&		&		& 	\yFb	&	\yF	&		&			&	0	&	$1$	&	0	\\ 
$\P_1$	&		&		&		&		&	\yFb	&	\yF	&			&	0	&	$-1$	&	0	\\ 
$\P_2$	&		&		&		&		&		&	\yFb	&	\yF		&	0	&	$1$	&	0	\\ \hline
\end{tabular}
\caption{Above, the original s-confining theory $A + 4 (Q_0 + \P_0)$ is extended on the left and right by gauging $G_L^2 \times \bG_R^2$ and adding the $Q_i$ and $\P_i$ fields to cancel the anomalies. To extend the model beyond $\ell= \arr=2$, more quarks $Q_i$ and $\P_j$ can be added with alternating $U(1)_A$ and $U(1)_B$ charges.
 }\label{table:UVmatterl2r2}
\end{table}

After extending the model in this way, the model has a ``left-right" symmetry which simplifies many of the calculations in this section:
\begin{align}
\ell \leftrightarrow \arr &,&
G_i \leftrightarrow \bG_i &,&
\Lambda_i \leftrightarrow \bL_i &,&
SU(4)_L \leftrightarrow SU(4)_R &,&
U(1)_A \leftrightarrow U(1)_B &,&
Q_i \leftrightarrow \P_i .
\end{align}
Above, $\Lambda_i$ corresponds to the group $G_i$, while $\bL_i$ is the confinement scale of the group $\bG_i$.
The group $G_0\times U(1)_R$ and the field $A$ are invariant under the discrete transformation.  %For example, after finding anomaly coefficients for $SU(4)_L$ and $U(1)_A$, we can apply the left-right transformation to find the analogous $SU(4)_R$ and $U(1)_B$ results.

\paragraph{Infrared operators:}

Based on our understanding of the $(\ell=0, \arr=k-1)$ models developed in the previous section and the vectorlike nature of the $G_0$-charged fields, we can guess the form of the gauge-invariant operators which describe the moduli space:
\begin{align}
\ff &\equiv \left\{
\begin{array}{rcl}
U_1 &=& A^2 \\
Z_i &=& Q_i^4 \\
\barZ_j &=& \P_j^4 ,
\end{array}
~~~~~~~
\begin{array}{rcl}
M_{\ell\arr} &=& (Q_\ell \ldots Q_1 Q_0 \P_0 \P_1 \ldots \P_\arr) \\
K_\ell &=& (Q_\ell^2 \ldots Q_0^2 A) \\
\barK_\arr &=& (A \P_0^2 \ldots \P_\arr^2) 
\end{array}
\right\} , \label{eq:su4ops}
\end{align}
for $i=0,1, \ldots, \ell$ and $j=0,1, \ldots, \arr$.

Only under certain conditions do we expect the basis $\ff$ to obey the anomaly matching conditions for the family symmetries listed in Table~\ref{table:UVmatterl2r2}. We have already seen that in the $(\ell=0,\arr=k-1)$ models with even $k$, 
some of the operators in $\ff$ become redundant in the $U(1)_B$ preserving vacuum.
%Based on the left-right symmetry for $N=4$, the same must be true for $U(1)_A$ in the $\ell=k-1$, $\arr=0$ models.
If this pattern continues in the $(\ell, \arr)$ models with $\ell\neq0$ and $\arr \neq 0$, then we would expect that the set $\ff$ obeys the anomaly matching conditions only if $\ell$ and $\arr$ are even. If either $\ell$ or $\arr$ is odd, we expect that some operators in $\ff$ become redundant if $U(1)_A\times U(1)_B$ is preserved in the vacuum.
%We therefore make the following ansatzes:
%\begin{itemize}
%\item If $\ell$ and $\arr$ are both even, then the set $\ff$ obeys the anomaly matching conditions for all of the exact symmetries;
%\item If $\ell$ or $\arr$ is odd, then the anomaly coefficients involving $U(1)_A$ or $U(1)_B$ match only after removing some operators.
%\end{itemize}
%As before, there are two options if $\ff$ does not satisfy the anomaly matching conditions: either the symmetry $U(1)_{A,B}$ is broken, or some of the operators in $\ff$ are redundant in the vacuum.

For a given $(\ell, \arr)$, the number of infrared operators is given by 
\begin{equation}
\dim \ff = 1+ (\ell+1) + (\arr+1) + 4^2 + \frac{4(3)}{2} + \frac{4(3)}{2} = \ell + \arr + 31 ,
\end{equation}
while the dimension of the classical moduli space is
\begin{equation}
\dim M_0 = (\ell+1) 4^2 + \frac{4(3)}{2} + (\arr+1) 4^2 - (\ell + 1 + \arr)(4^2 -1) = \ell + \arr + 23.
\end{equation}
This implies that there should exist $N_\text{con} =8$ constraint equations.

\paragraph{Equations of Motion:}

It is easiest to derive the equations of motion in the case where $G_0$ confines last. The groups $G_1 \times \ldots \times G_\ell$ and $\bG_1 \times \ldots \times \bG_r$ confine separately to form the mesons $M_L = (Q_0 \ldots Q_\ell)$ and $M_R = (\P_0 \ldots \P_\arr)$, the baryons $Z_{i=0 \ldots \ell}$ and $\barZ_{j=0\ldots \arr}$, and some larger baryon operators with quantum-modified constraints. The charges of $M_L$ and $M_R$ are shown in Table~\ref{table:UVmatterL0R}.
In the limit where $\Lambda_0$ is small, the theory reduces to two copies of $F=N$ \susy\ QCD with product group extensions. According to~\cite{Chang:2002qt}, the fields obey the following constraints:
\begin{eqnarray}
\det M_L &=& (Z_0 Z_1 \ldots Z_\ell) - \Lb{1} (Z_2 \ldots Z_\ell) - \ldots - (Z_0 \ldots Z_{\ell-2} ) \Lb{\ell} + \ldots \\
\det M_R &=& (\bZ_0 \bZ_1 \ldots \bZ_\arr) - \bLb{1} (\bZ_2 \ldots \bZ_\arr) - \ldots - (\bZ_0 \ldots \bZ_{\arr-2} ) \bLb{\arr} + \ldots
\end{eqnarray}
If $\ell$ is odd-valued, then the sum of neighbor contractions includes a constant term, $(\Lb{1}\Lb{3} \ldots \Lb{\ell})$; if $\ell$ is even, then all terms include some power of $Z_i$. The same relationship holds for $\arr$ and $\det M_R$. As in the $SU(N)^k$ models, we expect that the distinction between even and odd $\ell$ and $\arr$ determines which of the operators in $\ff$ are redundant when $U(1)_A$ and $U(1)_B$ are conserved in the vacuum.

\begin{table}[h]
\centering
\begin{tabular}{| c | c | c  | c | c c c |} \hline
   	& $SU(4)_L$	& $G_0$ 	& $SU(4)_R$	&$U_A$	&$U_B$	&$U_R$	\\ \hline
$M_L$	& \yFb	&	\yF	&			&$\{0,1\}$	&	$0$	&	0	\\
$A$		& 		& 	\yA	&			&	$-2$	&	$-2$	&	1	\\
$M_R$	& 		&	\yFb	&	\yF		&	$0$	& $\{0,1\}$	&	0	\\ \hline
\end{tabular}
\caption{All gauge groups except $G_0$ have confined, leaving $M_L$ and $M_R$. The $\{0,1\}$ charges of $M_L$ and $M_R$ correspond to the cases where $\ell$ and $\arr$ are odd or even, respectively. Not shown are the baryons $Z_i$ and $\bZ_j$, which do not transform under the non-Abelian symmetries.
}\label{table:UVmatterL0R}
\end{table}

When $G_0$ confines, $\{M_L, A, M_R\}$ form the following hadrons:
\begin{align}
\begin{array}{rcl}
U_1 &=& A^2 \\
Z_L &=& \det M_L \\
Z_R &=& \det M_R 
\end{array}
&&
\begin{array}{rcl}
M_{\ell\arr} &=& (M_L M_R) \\
K_\ell &=& (A M_L^2) \\
\barK_\arr &=& (A M_R^2) ,
\end{array}
\end{align}
with the dynamically-generated superpotential
\begin{equation}
W_d \sim \frac{A^2 \tM_L^4 \tM_R^4}{\tLb{0}} \sim \frac{U_1 Z_L Z_R - Z_R K_\ell^2 - Z_L \barK_\arr^2 - U_1 M_{\ell\arr}^4 + K_\ell M_{\ell\arr}^2 \barK_\arr }{\tLb{0} (\Lambda_1 \ldots \Lambda_\ell)^4(\bL_1 \ldots \bL_\arr)^4},
\end{equation}
for some $\tLb{0}$ consistent with the anomalous symmetries. We show the charges of the composite fields in Table~\ref{table:SU4LRfinal}.

%where
%\begin{equation}
%\tLb{0} \sim \Lb{0} \frac{\Lambda_\ell^4 \ldots \Lambda_1^4 \bL_1^4 \ldots \bL_\arr^4 }{(Z_\ell \ldots Z_0)(\bZ_0 \ldots \bZ_\arr)}.
%\end{equation}
The equations of motion from $U_1$, $K_\ell$, and $Z_L$ produce the following constraints:
\begin{align}
\begin{array}{rcl}
\det M_{\ell\arr} &=& Z_L Z_R  \\ 
U_1 M^3 &=& K_\ell M \barK_\arr ,\end{array}
&&
\begin{array}{rcl}
K_\ell Z_R &=& M_{\ell\arr}^2 \barK_\arr \\
\barK_\arr Z_L &=& K_\ell M_{\ell\arr}^2 ,\end{array}
&&
\begin{array}{rcl}
\Pf \barK_\arr &=& U_1 Z_R  \\
\Pf K_\ell &=& U_1 Z_L .  \end{array}
\label{eq:eomLR}
\end{align}
These equations are not all independent, but contain $N_\text{cons} = 8$ independent constraints.

\begin{table}[h]
\centering
\begin{tabular}{| c | c  c | c c | c c | c |} \hline
   	& $SU(4)_L$	& $SU(4)_R$	&$U_A^{\text{odd }\ell}$&$U_A^{\text{even }\ell}$&$U_B^{\text{odd }\arr}$	&$U_B^{\text{even }\arr}$	&$U_R$	\\ \hline
$K_\ell$	& \yA	&			&	$-2$			&		0			&	$-2$				&	$-2$				&	1	\\
$M_{\ell\arr}$& 	\yFb	&	\yF		&	$0$			&		1			&	$0$				&	$1$				&	0	\\
$\barK_\arr$& 		&	\yA		&	$-2$			&		$-2$			&	$-2$				&	$0$				&	1	\\ \hline
$U_1$	&		&			& $-4$			&		 $-4$			&	 $-4$				& 	$-4$				&	2	\\ 
$Z_{\text{even }i}$ &	&			& $+4$			&		 $+4$		&	 0				& 	0				&	0	\\
$Z_{\text{odd } i}$ &	&			& $-4$			&		 $-4$			&	 0				& 	0				&	0	\\ 
$\bZ_{\text{even }j}$ &	&		& $0$			&		 $0$			&	$+4$				& 	$+4$				&	0	\\
$\bZ_{\text{odd }j}$ &	&		& $0$			&		 $0$			&	$-4$				& 	$-4$				&	0	\\ \hline
\end{tabular}
\caption{After all of the gauge groups confine, the infrared degrees of freedom are described by the hadrons shown above. Their $U(1)_A$ and $U(1)_B$ charges depend on $\ell$ and $\arr$, respectively. }
\label{table:SU4LRfinal}
\end{table}

If we introduce Lagrange superfields $\lambda_L$ and $\lambda_R$,  the quantum modified constraints relating $\{Z_L, Z_R\}$ to $\{Z_i, \bZ_j\}$ as a superpotential:
\begin{eqnarray}
W_L &=& \lambda_L \Big( Z_L - (Z_0 Z_1 \ldots Z_\ell) + \Lb{1} (Z_2 \ldots Z_\ell) + \ldots + (Z_0 \ldots Z_{\ell-2} ) \Lb{\ell} + \ldots \Big) \\
W_R &=& \lambda_R \left( Z_R -  (\bZ_0 \bZ_1 \ldots \bZ_\arr) + \bLb{1} (\bZ_2 \ldots \bZ_\arr) + \ldots + (\bZ_0 \ldots \bZ_{\arr-2} ) \bLb{\arr} + \ldots \right).
\end{eqnarray}
%The equations of motion set
%\begin{align}
%\lambda_L = - \frac{U_1 Z_R - \barK_\arr^2}{\tLb{0}  (\Lambda_1 \ldots \Lambda_\ell)^4(\bL_1 \ldots \bL_\arr)^4} &&
%\lambda_R = - \frac{U_1 Z_L - K_\ell^2}{\tLb{0}  (\Lambda_1 \ldots \Lambda_\ell)^4(\bL_1 \ldots \bL_\arr)^4},
%\end{align}
%so that $\lambda_{L}$, $\lambda_R$, $Z_L$ and $Z_R$ are all redundant operators.

\paragraph{Redundant Operators:}

In this section we use the equations of motion to study the operator basis $\ff$. 
%The phases of the $\Lb{i}$ and $\bLb{j}$ are not important in this section, so we ignore them. 
In the $U(1)_A $ preserving vacuum with $\ev{Z_i} = 0$, the expectation value of $Z_L$ depends heavily on whether $\ell$ is even or odd. If $\ell$ is even, then $Z_L\approx 0$;  if $\ell$ is odd, then $Z_L \approx (\Lambda_1^b \Lambda_3^b \ldots \Lambda^b_\ell) \gg 0$. The same pattern holds for $\arr$ and $\bZ_j$ when $U(1)_B$ is preserved. 

It is simplest to consider the case in which both $\ell$ and $\arr$ are even. Expanding about the $Z_i = \bZ_j = 0$ vacuum to first order in $Z_i$ and $\bZ_j$, we find that every term in \eqref{eq:eomLR} contains a product of at least two fields, so that none of the operators in the set $\ff$ are redundant. 
This is consistent with the fact that all of the anomaly coefficients from $SU(4)_L \times SU(4)_R \times U(1)_A \times U(1)_B \times U(1)_R$ match the ultraviolet theory when $\arr$ and $\ell$ are even. %, including the mixed gravitational-$U(1)$ anomalies.
%Therefore, the operator set $\ff$ describes the degrees of freedom near the origin of moduli space.

This is not true if $\ell$ is odd. In this case the equations of motion for $\barK_\arr Z_L$ and $U_1 Z_L$ can be rewritten as
\begin{align}
\barK_\arr = \frac{K_\ell M_{\ell\arr}^2 }{(\Lb{1} \Lb{3} \ldots \Lambda^b_\ell)}  ,&&
U_1 = \frac{\Pf K_\ell}{(\Lb{1} \Lb{3} \ldots \Lambda^b_\ell)} . \label{eq:ellodda}
\end{align}
near the $U(1)_A\times U(1)_B$ preserving vacuum. Similarly, the equation of motion for $\det M_{\ell\arr}$ becomes
\begin{equation}
\bZ_0 (\bLb{2} \bLb{4} \ldots \bLb{\arr} ) + \bLb{1} \bZ_2 (\bLb{4} \ldots \bLb{\arr} ) + \ldots + (\bLb{1} \bLb{3} \ldots \bLb{\arr-1} ) \bZ_\arr = \frac{ \det M_{\ell\arr} }{(\Lb{1} \Lb{3} \ldots \Lb{\ell} )} ,  \label{eq:elloddb}
\end{equation}
which can be recast into a linear constraint equation for any one of the $\bZ_\text{even}$ fields. Taken together, Eqs.~(\ref{eq:ellodda}) and~(\ref{eq:elloddb}) imply that the operators $\{\barK_\arr, U_1, \bZ_\text{even} \}$ should be removed in the $U(1)_A \times U(1)_B$ preserving vacuum if $\ell$ is odd and $\arr$ is even.
%It can be seen that when these operators are removed, the anomaly coefficients match again.
%
In the even~$\ell$, odd $\arr$ case it is the operators $\{K_\ell, U_1, Z_\text{even} \}$ which become redundant, and $Z_R$ rather than $Z_L$ remains large in the $\bZ_j = 0$ vacuum.

If both $\ell$ and $\arr$ are odd, then the origin of moduli space is no longer a solution to the equations of motion: %This is due to the equations of motion for $\det M$:
\begin{eqnarray}
 \det M_{\ell\arr}  &=& (\Lb{1}\Lb{3} \ldots \Lb{\ell} )(\bLb{1} \bLb{3} \ldots \bLb{\arr} ) - \Big( Z_0 Z_1 \Lb{3} \ldots \Lb{\ell} + Z_0 \Lb{2}  Z_3 \ldots \Lb{\ell} + \ldots \Big)(\bLb{1}  \ldots \bLb{\arr} ) \nonumber\\&&\ - (\Lb{1} \ldots \Lb{\ell} )\left( \bZ_0 \bZ_1 \bLb{3} \ldots \bLb{\arr} + \bZ_0 \bLb{2}  \bZ_3 \ldots \bLb{\arr} + \ldots \right) + \ldots
\end{eqnarray}
To satisfy this constraint, either $\ev{M}\neq 0$, $\ev{\Zeven \Zodd} \neq0$, or $\ev{\bZeven \bZodd} \neq 0$. Different family symmetries are broken in each case, leaving different sets of independent operators. 

In the $\ev{M}\neq 0$ vacuum where $M^i_j$ is proportional to $\delta^i_j$, $SU(4)_L \times SU(4)_R$ is broken to its diagonal subgroup $SU(4)_d$. The fields $Q_\ell$ and $\barQ_\arr$ transform under $SU(4)_d$ as $\yFb$ and $\yF$, respectively, while the meson $M$ decomposes as
\begin{align}
\yFb \otimes \yF = \bf 1 \oplus \yAd : && M_{\ell\arr} \longrightarrow ( \Tr M_{\ell\arr} ) \oplus (M_{\ell\arr} - \Tr M_{\ell\arr} ) .
\end{align}
In the $U(1)_A \times U(1)_B$ preserving vacuum with $Z_i = \bZ_j =0$, it is possible to write $\barK_\arr$ and $U_1$ either in terms of $K_\ell$ and $M_{\ell\arr}$, or $K_\ell$ and $U_1$ in terms of $\barK_\arr$ and $M_{\ell\arr}$. Therefore, we can either remove the set $\{ K_\ell, U_1, \Tr M\}$ or $\{\barK_\arr, U_1, \Tr M\}$. This degeneracy is related to the fact that $K_\ell$ and $\barK_\arr$ have the same transformation properties under $SU(4)_d \times U(1)_A \times U(1)_B \times U(1)_R$.
%After removing these operators, the anomaly coefficients match.

If instead $\ev{M}=0$ and $\ev{\Zeven \Zodd} \neq0$, only $U(1)_A$ is broken in the vacuum. One ``$(\Zeven + \Zodd)$" linear combination determined by the ratio of the expectation values becomes massive, and all sixteen $M^i_j$ degrees of freedom remain independent. %The precise linear combination of $Z_i$ appearing in $(\Zeven + \Zodd)$ depends on the ratios of the vacuum expectation values. 
The operator $\barK_\arr$ is not redundant in this vacuum: the $Z_L \barK_\arr$ equation of motion includes a term $\Zeven \Zodd \barK_\arr$ which is not small.
The set of redundant operators is $\{K_\ell, U_1, (\Zeven + \Zodd)\}$.

Finally, if the nonzero expectation value is $\ev{\bZeven \bZodd}$, then $U(1)_B$ is broken. As we would expect from the left-right symmetry, the redundant operators are $\{\barK_\arr, U_1,  (\bZeven + \bZodd) \}$ in this vacuum.
It is also possible to break a linear combination of $U(1)_A$ and $U(1)_B$ if $\ev{\Zeven \Zodd}\neq 0$ and $\ev{\bZeven \bZodd}\neq0$.
%Particularly interesting combinations include $U(1)_A \pm U(1)_B$, which are symmetric and antisymmetric (respectively) under the left-right symmetry.

\paragraph{Anomaly Matching:}

We have discussed six distinct cases with maximal symmetry in the vacuum, based on $\ell$ and $\arr$. Below, we show a summary of our results for each case:
\begin{center}
\begin{tabular}{|c | c | l  |} \hline
 $(\ell,\arr)$   & Broken symmetry & Redundant operators \\ \hline
(even, even) 	& None	&  None \\
(odd, even)		& None	& $\{ \barK_\arr, U_1, \bZeven \}$ \\
(even, odd) 		& None	& $\{K_\ell, U_1, \Zeven\}$ \\ \hline
		& $SU(4)_L \times SU(4)_R$ & $\{K_\ell \text{ or } \barK_\arr, U_1, \Tr M_{\ell\arr}  \}$ \\
(odd, odd) 	& $U(1)_A$ & $\{ K_\ell, U_1, (\Zeven+\Zodd) \}$ \\
		&  $U(1)_B$ & $\{ \barK_\arr, U_1, (\bZeven + \bZodd) \}$ \\ \hline
\end{tabular}
\end{center}
For the remaining symmetries and operators in each case, we have verified that the anomaly coefficients match the UV theory. There are 21 matching conditions for each of the first three cases, 17 for the fourth case, and 12 each for the final two cases. Although some of these coefficients are related to each other via the left-right symmetry, the explicit calculation is lengthy and not very illuminating.

Let us also consider points on the moduli space with nonzero $\ev{Z_i}$ or $\ev{\bZ_j}$, where none of the operators in the set $\ff$ are redundant. In these vacua $U(1)_A \times U(1)_B$ is spontaneously broken, and the infrared operators should obey anomaly matching conditions for the remaining symmetries.

For the odd $\ell$, even $\arr$ case, $U(1)_A$ is broken by $\ev{Z_i}\neq 0$ for some $Z_i$.
After $U(1)_A$ is broken, $\{U_1, \bZeven\}$ form an anomaly-neutral pair: their $U(1)_{B,R}$ charges are opposite, so all of the $U(1)^3$ and gravitational $U(1)$ anomalies cancel.
The fermionic part of $\barK_\arr$ is neutral under $U(1)_B \times U(1)_R$, and it is in a real representation of $SU(4)_R$: therefore, $\barK_\arr$ contributes nothing to the remaining anomaly coefficients. 
Thus, the t'~Hooft anomaly matching conditions are also satisfied in the $\ev{Z_i}\neq0$ vacuum where the operators $\{ \barK_\arr, U_1, \bZeven \}$ are independent degrees of freedom.

In the even-$\ell$, odd-$\arr$ models, the operators $\{K_\ell, U_1, \Zeven\}$ are restored as independent degrees of freedom when $\ev{\bZ_j}\neq0$ and $U(1)_B$ is spontaneously broken.
Applying the left-right transformation to the above results, the introduction of $\{K_\ell, U_1, \Zeven\}$ has no net effect on the anomaly coefficients once $U(1)_B$ is removed.
Finally, when $\ev{Z_i}\neq 0$ and $\ev{\bZ_j}\neq 0$ in the odd-$\ell$, odd-$\arr$ models, the operators $\{K_\ell, U_1, \Zeven\}$ are restored as independent degrees of freedom without contributing to the anomaly coefficients of the remaining symmetries. Both $U(1)_A$ and $U(1)_B$ are broken in this case.

\paragraph{Flows:}

Our proposed s-confining extensions to the $SU(4)$ model pass several consistency checks. As a final test, let us spontaneously break the gauge group by giving large expectation values to the gauge invariant operators, as in Section~\ref{sec:flow}.
For example, $\ev{\mlr} \gg \Lambda$ breaks $SU(4)^{\ell+\arr+1}$ to $SU(3)^{\ell+\arr+1}$, leaving $\yA+4 \yF + 3 \yFb $ matter charged under $SU(3)_0$. Three of the $\yF$ fields come from the $G_0 \times G_1$ bifundamental $Q_0$, while the fourth comes from 
\begin{align}
SU(4)\rightarrow SU(3):~~~~~
{\yA} \longrightarrow \yA \oplus \yF \, .
\end{align}
Note that $\yA = \yFb$ for $SU(3)$, so that there are effectively $(3+1)$ flavors of $(\yF + \yFb)$ charged under $SU(3)_0$. 
%This is the s-confining $SU(3)$ \susy\ QCD model, with a product group extension.
The low-energy theory is a left-right extension of $F=4$, $N=3$ \susy\ QCD, where an $SU(3)_L\times SU(3)_R$ subgroup of the family $SU(4)_L \times SU(4)_R$ is gauged. In Section~\ref{sec:sqcdp} we consider such models in more detail.

Along flat directions with $\ev{\Pf A}\gg \Lambda_0$, $SU(4)_0$ is broken to $Sp(4)$, leaving an $(\ell, \arr)$ product group extension of the s-confining $Sp(4): (4+4) \yF$ model. In this theory an $SU(4)_L\times SU(4)_R$ subgroup of the $SU(8)$ family symmetry is gauged.
We discuss models of this type in Section~\ref{sec:spinextension}. %, along with the product group \susy\ QCD model that follows from $\ev{\mlr}$.

\paragraph{Summary:}

In every $(\ell,\arr)$ model with $(\ell,\arr)\neq(0,0)$, there are quantum deformations to the classical moduli space. The origin remains on the moduli space unless both $\ell$ and $\arr$ are odd. In the mixed case where only one of $\{\ell, \arr\}$ is odd, eight of the fields become redundant in the vacua which conserve $U(1)_A\times U(1)_B$. 
%So, if $\ell$ or $\arr$ are odd, then in the symmetry-enhanced vacua---which includes the origin of moduli space if the sum $(\ell+\arr)$ is odd---eight of the fields become redundant, and $W_d$ gives $\mathcal O(\Lambda)$ masses to their independent degrees of freedom. The remaining $(\dim \ff - 8)=\dim M_0$ degrees of freedom are effectively free. Away from the symmetry-enhanced point on $M_0$, the operators $\ff$ are independent, interacting degrees of freedom. 
If $\ell$ and $\arr$ are both even, all of the infrared operators in \eqref{eq:su4ops} are independent, interacting degrees of freedom even at the origin of moduli space.
%If at least one of $\{\ell, \arr\}$ is even, then it is possible to have confinement without chiral symmetry breaking, as in s-confining models.
Due to the existence of a dynamically generated superpotential and the possibility of confinement without chiral symmetry breaking, we conclude that the $(\ell,\arr)$ models are s-confining if $\ell$ and $\arr$ are not both odd.

\paragraph{$SU(4)$ Ring Extension:}

Before moving on to consider other types of models, let us extend the $(\ell, \arr)$ model even further by gauging a diagonal subgroup $G_d$ of the family $SU(4)_L \times SU(4)_R$ symmetry.
This connects the left and right ends 
%Under this new $G_d$, the fields $Q_\ell$ and $\P_\arr$ transform as $\yFb$ and $\yF$, respectively, thus connecting the left and right ends 
 of the $(\ell, \arr)$ extension as shown in Table~\ref{table:ringSU4}, % we show an $(\ell\neq0,\arr=0)$ model: other arrangements can be obtained by rotating the labels on the groups $G_i$. %Of the original $\ell+2$ family $U(1)$ symmetries of the scalar fields, $\ell+1$ of these are broken by the gauge anomalies. There is also an unbroken $U(1)_R$ symmetry, under which $A$ has charge $+1$.
 so that different models are labelled by the sum $(\ell+\arr)$.
Models of this type appear in deconstructions of 5d gauge theories, as in~\cite{Csaki:2001em}. 

\begin{table}[h]
\centering
\begin{tabular}{| c | c c c c c |} \hline
   		& $G_\ell$	&$G_{\ell-1}$& $\ldots$	& $G_1$	& $G_0$	\\ \hline
$Q_\ell$	& \yF		&		&		&		&	\yFb		\\
$Q_{\ell-1}$&	\yFb	& \yF		&		&		&			\\
$\vdots$	&		& \yFb	&		&		&			\\
$Q_1$	&		& 		&$\ddots$	&	\yF	&			\\ 
$Q_0$	&		& 		&		&	\yFb	&	\yF		\\ 
$A$		&		& 		&		&		&	\yA		\\ \hline 
\end{tabular}
\caption{Above, we show the matter fields of the $SU(4)$ ring extension to the $A+4Q + 4 \barQ$ model. %It has a Coulomb branch, and is therefore not s-confining.
 }\label{table:ringSU4}
\end{table}

Although the baryon operators $\Pf A$ and $\det Q_i$ are unaffected by the ringlike nature of the product gauge group, there is now only one gauge-invariant meson operator: $\Tr M = \Tr (Q_0 Q_1 \ldots Q_\ell)$.
For any group $G_i$, the adjoint operator
\begin{equation}
(\hat{M_i})^\alpha_\beta  = (Q_i Q_{i+1} \ldots Q_\ell Q_0 \ldots Q_{i-1} )^\alpha_\beta -\frac{1}{4} (\Tr M) \delta^\alpha_\beta 
\end{equation}
is a degree of freedom in the limit where $G_i$ is weakly gauged, and can be used to create gauge-invariant operators of the type $\Tr (\hat M_i \hat M_i)$ and $\Tr (\hat M_i^3)$. In this notation, $Q_{-1} = Q_\ell$ for the $i=0$ case.

Even when these operators have large expectation values, the gauge group is not completely broken. It has been shown~\cite{Csaki:1997zg} in the $SU(N)^k$ extension to $F=N$ \susy\ QCD that at an arbitrary point on the moduli space has a remaining $U(1)^3$ gauge group.
In the $A+4Q + 4\barQ$ model it is also possible to set $\ev{\Pf A}\gg \Lambda_0$, so that $SU(4)_0$ is broken to $Sp(4)$. This reduces the rank of the group by one, but is not sufficient to break $U(1)^3$ completely. 
Therefore, the $SU(4)$ ring extension has a Coulomb branch, and is not s-confining.

\subsection{$Sp(2m)$ with $(2m+4)$ quarks} \label{sec:spinextension}

In Section~\ref{sec:disorderedWd}, we found that the $SU(N)^k$ extension of the $A+4Q+N\barQ$ model flows to an $Sp(2m)\times SU(2m)^{k-1}$ theory.
In the limit where $Sp(2m)$ is much more strongly coupled than the $SU(2m)$ groups, 
the $(2m+4)$ quarks confine to produce the operator $M=(Q^2)$, which transforms in the $\yA$ representation under the approximate $SU(2m+4)$ family symmetry.

The fields $Q$ and $M$ have the following charges:
\begin{center}
\begin{tabular}{|r | c | c c |} \hline
	& $Sp(2m)$	& $SU(2m+4)$	& $U(1)_R$ \\ \hline
$Q$	&	\yF		&	\yF		& $1/(m+2)$ \\ \hline\hline
$M$ &		& 	\yA		& $2/(m+2)$ \\ \hline
\end{tabular}
\end{center}
A dynamically generated superpotential
\begin{equation}
W_d = \frac{\Pf M}{\Lambda^{2m+1}}
\end{equation}
reproduces the classical constraints on the $Q_i$ fields.

\begin{table}[h]
\centering
\begin{tabular}{| c | c | c | c c c | c |} \hline
   	& $SU(4)_L$	&$Sp(2m)$&$SU(2m)_1$& $\ldots$	& $SU(2m)_k$	& $SU(2m)_R$\\ \hline
$Q_L$	& \yF		&	\yF	&		&			&		&	\\
$\P_0$	&		& \yF		&	\yF	&			&		&	\\
$\P_1$	&		& 		&	\yFb	&			&		&	\\
$\vdots$	&		& 		&		&	$\ddots$	&		&	\\ 
$\P_{k-1}$	&		& 		&		&			&	\yF		&	\\ 
$\P_k$	&		& 		&		&			&	\yFb		& \yF 	\\ \hline \hline
$(Q_L^2)$ & \yA	& 		&		&			&			&		\\
$(Q_L \P_0)$ & \yF	& 		& 	\yF	&			&			&		\\	
$(\P_0^2)$&		& 		&	\yA	&			&			&		\\ 	
$\P_1$	&		& 		&	\yFb	&			&		&	\\
$\vdots$	&		& 		&		&	$\ddots$	&		&	\\ 
$\P_k$	&		& 		&		&			&	\yFb		& \yF 	\\ \hline 
\end{tabular}
\caption{An $Sp(2m) \times SU(2m)^k$ model is shown, which is expected to s-confine. At the bottom of the table, we list the degrees of freedom in the confined phase of $Sp(2m)$. Subsequent confinement follows the pattern of the $A+4Q + N\barQ$ model. }\label{table:UVmatterSpN}
\end{table}

In the product gauge group model shown in Table~\ref{table:UVmatterSpN}, an $SU(2m)$ subgroup of the family symmetry is gauged and new bifundamental fields are added to cancel the anomalies. The family $SU(2m+4)$ is explicitly broken to $SU(2m) \times SU(4) \times U(1)$, under which the meson $M$ decomposes as 
\begin{align}
\yA  \longrightarrow (\yA,  {\yI}; -4) \oplus (\yF, \yF; m-2) \oplus ({\yI}, \yA; 2m) &:&
M  \longrightarrow M_A \oplus M_Q \oplus M_0 ,
\end{align}
%
%\begin{center}
%\begin{tabular}{r | c c c |}
%	& $SU(2m)$	& $SU(4)$	& $U(1)$ \\ \hline
%$M_A$&	\yA		&		& $-4$ \\ 
%$M_Q$ &	\yF	& 	\yF		& $m-2$ \\ 
%$M_0$ &		&	\yA		& $2m$	\\ \hline
%\end{tabular}
%\end{center}
and the dynamically generated superpotential becomes
\begin{equation}
W_d \longrightarrow \frac{M_A^{m-1} M_Q^2 M_0 }{\Lambda^{2m+1}} .
\end{equation}
Including the bifundamental field $\P_1$, the $SU(2m)_1$ charged matter in the confined phase of $Sp(2m)$ is $M_A + 4 M_Q + 2m \barQ_1$, which is expected to s-confine.

%As discussed in Section~\ref{sec:flow}, the $Sp(2m)\times SU(2m)^{k-1}$ theory can be derived from the $A+4Q + 2m\barQ$ $SU(2m)^k$ model, in the limit $\ev{\Pf A} \gg \Lambda$. By integrating out $\Pf A = U_1$ from the $SU(N)^k$ theory, one can derive the dynamically generated superpotential for the $Sp(2m)$ model.
%In this way, we can derive the dynamically generated superpotential from the $SU(2m)$ result, \eqref{eq:fullWd}.

This model can also be derived using the deconfinement technique of Berkooz~\cite{Berkooz:1995km}, by treating the matter field $A$ as a bound state of two quarks transforming in the fundamental representation of a new $Sp(N)$.

\subsection{SUSY QCD}\label{sec:sqcdp}

%Rather like $Sp(2m)$ with $(2m+4)$ quarks, 
A product group extension to $F= N+1$ \susy\ QCD can be derived from the $N=3$ case of $A+4 Q + N\barQ$. In $SU(3)$, the $\yA$ representation is the same as $\yFb$, so that the $G_1$ matter is effectively $4 \yF + 4 \yFb$. By gauging the $SU(3)$ family symmetry of the $\barQ$ and adding a sequence of bifundamental fields $\P_i$, we have found a product group extension to \susy\ QCD.

For larger values of $N$, let us gauge an $SU(N)$ subgroup of the $SU(N+1)_R$ family symmetry as shown below:
\begin{center}
\begin{tabular}{|r | c | c c | c |}\hline
& $SU(N+1)_L$	& $SU(N)_1$	& $SU(N)_2$	& $SU(N)_R$ \\ \hline
$Q$	&	\yF		&	\yF		& 			&			 \\ 
$\bar q$ &			& 	\yFb		&			&			 \\ 
$\P_1$ &			& 	\yFb		&	\yF		&			 \\ 
$\P_2$ &			& 			&	\yFb		&	\yF		 \\ \hline
\end{tabular}
\end{center}
After $SU(N)_1$ confinement, the hadrons are $(Q \bar q)$, $(Q \P_1)$, $(Q^N)$, $(\P_1^N)$, and $(\bar q \P_1^{N-1} )$, which transform under $SU(N)_2$ and the family symmetries as: 
%After confinement, the degrees of freedom are:
\begin{center}
\begin{tabular}{|r | c | c | c |}\hline
	& $SU(N+1)_L$	& $SU(N)_2$	& $SU(N)_R$ \\ \hline
$(Q\bar q)$&	\yF		& 			&			 \\ 
$(Q^N)$ &	\yFb		&			&			 \\ 
$(\P_1^N)$ &			&			&			 \\ \hline
$(Q\P_1)$ &	\yF		&	\yF		&			 \\ 
$(\bar q\P_1^{N-1})$ &	&	\yFb		&			 \\ 
$\P_2$ &				& 	\yFb		&	\yF		 \\ \hline
\end{tabular}
\end{center}
%Above the $SU(N)_2$ confinement scale,
Under $SU(N)_2$ there are $(N+1)(\yF + \yFb)$ matter fields, which is consistent with the index constraint for s-confinement.
%It is also possible to gauge a subgroup of $SU(N+1)_L$, as in the $(\ell, \arr)$ $A+4Q + 4\barQ$ model.

For this theory to be s-confining, it must be shown that the dynamically generated superpotential from $SU(N)_1$ does not prevent the operators $(Q\P_1)$ and $(\bar q \P_1^{N-1})$ from varying independently; that the infrared operators obey the appropriate anomaly matching conditions; and that the origin is on the moduli space.
The additional gauge groups are likely to introduce quantum-modified constraints between some of the operators, which may induce chiral symmetry breaking in some cases.

This theory can also be extended by gauging an $SU(N)$ subgroup of $SU(N+1)_L$, so that the most general product group extension is $SU(N)^\ell \times SU(N)_0 \times SU(N)^\arr$. Based on the behavior of the $(\ell, \arr)$ $A+4Q + 4\barQ$ model for odd $\ell$ and $\arr$, we expect that some of the $(\ell,\arr)$ \susy\ QCD models also break chiral symmetry.

%As in the $SU(4)$ $A+ 4Q + 4\barQ$ model, \susy\ QCD can be extended on the left by gauging $SU(N) < SU(N+1)_L$, with $Q\rightarrow  Q_1 \oplus q =\yF \oplus \bf 1 $.
%We encounter the $N=3$ version of this model in Section~\ref{sec:su4LR}, when the operator $\mlr$ acquires a large VEV and breaks $SU(4)^{\ell+\arr+1}$ to $SU(3)^{\ell+\arr+1}$.

\paragraph{Alternating Gauge Groups:}

The $F=N+1$ model can also be extended by gauging the entire $SU(N+1)$ family symmetry. In this case, the gauge group has the alternating form $SU(N)\times SU(N+1) \times SU(N) \times SU(N+1) \times \ldots$, with a series of bifundamental fields:
\begin{center}
\begin{tabular}{|r | c | c c c | c |}\hline
& $SU(N+1)_L$	& $SU(N)_1$	& $SU(N+1)_2$& $SU(N)_3$ 	& $SU(N+1)_R$\\ \hline
$Q$	&	\yFb		&	\yF		& 			&			&			 \\ 
$\P_1$ &			& 	\yFb		&	\yF		&			&			\\ 
$\P_2$ &			& 			&	\yFb		&	\yF		&			\\
$\P_3$ &			& 			&			&	\yFb		&	\yF		\\ \hline
\end{tabular}
\end{center}
The matter content is simpler in this case, as all of the fields are $SU(N+1) \times SU(N)$ bifundamentals. When $SU(N)_1$ confines, we are left with
\begin{center}
\begin{tabular}{|r | c | c c | c |}\hline
	& $SU(N+1)_L$	& $SU(N+1)_2$& $SU(N)_3$ 	& $SU(N+1)_R$\\ \hline
$(Q^N)$	&	\yF		& 			&			&			 \\ 
$(Q\P_1)$ &	\yFb		&	\yF		&			&			\\ 
$(\P_1^N)$ &			&	\yFb		&			&			\\ 
$\P_2$ &			&	\yFb		&	\yF		&			\\
$\P_3$ &			&			&	\yFb		&	\yF		\\ \hline
\end{tabular}
\end{center}
Under $SU(N+1)_2$, there are $(N+1)$ flavors of $\yF + \yFb$ which is expected to confine with chiral symmetry breaking. Many of the $G_2$ singlets we would na\"ively construct, such as $(Q \P_1)(\P_1^N)$, are set to zero by the equations of motion, so $G_2$ confinement leaves the following charged fields:
\begin{center}
\begin{tabular}{|r | c | c | c |}\hline
	& $SU(N+1)_L$	& $SU(N)_3$ 	& $SU(N+1)_R$\\ \hline
$(Q^N)$	&	\yF		&			&			 \\ 
$(Q\P_1\P_2)$ &	\yFb	&	\yF		&			\\ 
$\P_3$ &				&	\yFb		&	\yF		\\ \hline
\end{tabular}
\end{center}
After $G_1\times G_2$ confinement, the low energy theory is simply $F=N+1$ \susy\ QCD with some gauge singlet fields.

Both product group models based on \susy\ QCD have the potential to be s-confining, and may be promising directions for future study.

\subsection{Other Models}  \label{sec:notsconf}

Of the s-confining theories listed in~\cite{Csaki:1996zb}, there are only a few models possessing non-Abelian family symmetries larger than the gauge group. We have already discussed the $SU(N)$ models with $A+4Q+N\barQ$ and $(N+1)(Q + \barQ)$, as well as the $Sp(2m)$ model with $(2m+4)Q$. There are two remaining cases based on $Sp(2m)$ with $A+6Q$~\cite{Cho:1996bi,Csaki:1996eu}. If $m=2$ or $m=3$, an $SU(4)$ or $SU(6)$ subgroup of the family symmetry can be gauged. In this section, we show that the product group extensions do not exhibit s-confinement.

%\paragraph{$Sp(2m)$ with $A+6 Q$:}
%As a final example, we consider the s-confining $Sp(2m): \yA + 6 \yF$ models with $m=3$ and $m=2$. This exhausts the list of s-confining theories in~\cite{Csaki:1996zb} with family symmetries equal in size or larger than the confining gauge group. Unlike the previous examples, it does not appear that the typical product group extension produces s-confining theories in this case.

\paragraph{$Sp(6)$ with $A+6 Q$:}
Consider the $m=3$ case with just one extra product group. Below, we show the matter fields above and below the $Sp(6)$ confinement scale:
\begin{center}
\begin{tabular}{|r | c  c | c |}\hline
	& $Sp(6)$	& $SU(6)$		& $SU(6)_R$ \\ \hline
$A$	&	\yA	& 			&			 \\ 
$Q$ &	\yF	&	\yF		&			 \\ 
$\P$ &		&	\yFb		&	\yF		 \\ \hline\hline
$(A^2)$	&	& 			&			\\
$(A^3)$	&	& 			&			\\
$(Q^2)$	&	& 	\yA		&			\\
$(QAQ)$	&	& 	\yA		&			\\
$(QA^2Q)$&	& 	\yA		&			\\
$\P$ &		&	\yFb		&	\yF		 \\ \hline
\end{tabular} 
\end{center}
In the confined phase of $Sp(6)$, the $SU(6)$ index sum becomes
\begin{equation}
\sum_j \mu_j - \mu_G = 3 \cdot (6-2) + 6 \cdot 1 - 2 \cdot 6 = +6,
\end{equation}
so the product group does not s-confine. It may be possible to remove some of the degrees of freedom by adding a nonzero tree-level superpotential, but this is outside the scope of the current study.

\paragraph{$Sp(4)$ with $A+6 Q$:}

In the $Sp(4)$ case, an $SU(4)$ subgroup of the $SU(6)$ family symmetry is gauged. 
\begin{center}
\begin{tabular}{|r | c | c c | c |}\hline
   &$SU(2)_L$	& $Sp(4)$	& $SU(4)$		& $SU(6)_R$ \\ \hline
$Q_L$ & \yF	&	\yF	&			&			\\
$A$ &		&	\yA	& 			&			 \\ 
$Q_R$ &		&	\yF	&	\yF		&	 \\ 
$\P$ &		&		&	\yFb		&	\yF		 \\ \hline 
\end{tabular}
\end{center}
The set of $Sp(4)$ invariants is 
\begin{equation}
\ff = \{(A^2); (Q_L^2), (Q_L Q_R), (Q_R^2); (Q_L A Q_L), (Q_L A Q_R), (Q_R A Q_R) \}.
\end{equation}
The operators $(Q_L Q_R)$ and $(Q_L A Q_R)$ are bifundamentals of $SU(2) \times SU(4)$, while $(Q_R^2)$ and $(Q_R A Q_R)$ transform as $(\bf 1, \yA)$. The other hadrons are gauge singlets.
Together with $\P$, the $SU(4)$ charged matter is $2 \yA + 4 \yF + 4\yFb$, with the index sum
\begin{equation}
\sum_j \mu_j - \mu_G = 2 (2) + 4 (1) + 4 (1) - 2\cdot 4 = +4 .
\end{equation}
Therefore, the $Sp(4)$ product group extension to $Sp(4):(A+6Q)$ is also not s-confining.

%As suggested at the beginning of Section~\ref{sec:misc}, it may be possible to construct s-confining product group theories without using sequences of bifundamentals.

\section{Conclusion}

For several s-confining theories, we find product gauge group models with the following properties:
\begin{itemize}
\item All infrared degrees of freedom are gauge invariant composite fields;
\item The infrared physics is described by a smooth effective theory, which is valid everywhere on the moduli space (including the origin);
\item There is a dynamically generated superpotential.
\end{itemize}
This allows confinement without symmetry breaking, even when the quantum and classical moduli spaces are different. In particular, this behavior may be found in the following models:
\begin{align*}
SU(N): A+4Q + N\barQ &&
Sp(2m): (2m+4) Q &&
SU(N): (N+1)(Q+\barQ).
\end{align*}
In this paper we argue that the $A+4Q+N\barQ$ and $Sp(2m): (2m+4)Q$ product group models s-confine. Based on less rigorous arguments we suggest two product group extensions of \susy\ QCD which may also be s-confining, but a more detailed analysis is required.
It is also entirely possible that there are many other s-confining product group theories unrelated to the models considered in this paper.

In the $A+4Q+N\barQ$ model with $N=4$, we consider a set of product group extensions of the form $G_L^\ell \times G_0 \times G_R^\arr$. When $\ell$ and $\arr$ are both odd, the chiral symmetry is necessarily broken in the vacuum, so the theory is not s-confining. If instead the sum $(\ell + \arr)$ is odd, then the origin remains on the quantum-deformed moduli space, and some of the infrared operators become redundant in the symmetry-enhanced vacua. Finally, if $\ell$ and $\arr$ are both even, we find that all of the operators are interacting degrees of freedom in the neighborhood of the origin. 
In each case, there is a dynamically generated superpotential.

%By considering the flow of the $SU(N)$ $A+4Q+N\barQ$ theory, we have found an s-confining product group extension for $Sp(2m)$ with $(2m+4)$ quarks. We also propose \susy\ QCD product groups which appear likely to s-confine as an area for further study. 

%It may also be fruitful to analyze the $SU(N)^k$ product group models as being derived from a five-dimensional $\mathcal N=2$ \susy\ gauge theory.

%: it would be particularly useful to discover a systematic way to search for such theories.

%Our results for the s-confining product groups may be useful for composite model building. 
One feature of the product group models is the lack of small gauge-invariant operators, which has a promising phenomenological application to composite axion models. After lifting some of the flat directions, a Peccei-Quinn $U(1)$ symmetry may be dynamically broken when the gauge group confines, producing a light composite axion.
If the product gauge group is suitably large, the Peccei-Quinn symmetry is protected against the explicit symmetry breaking effects which would otherwise be induced by higher-dimensional operators.
We explore this option in an upcoming paper~\cite{compositeaxions}.

Another promising direction for future study is to treat the product gauge groups as $k$ site decompositions of 5d \susy\ theories. Exact calculations in $\mathcal N=2$ \susy\ may provide us with a better understanding of the 4d $\mathcal N=1$ models considered in this paper.

\section*{Acknowledgments}

We would like to thank Csaba Csaki, Yuri Shirman, Tim M.P. Tait, Flip Tanedo, and Arvind Rajaraman for helpful conversations. 
This work was supported in part by NSF Grants PHY-1316792 and PHY-1620638, and in part by Simons Investigator Award \#376204.

\appendix

\section{Derivation of classical constraints} \label{sec:appA}

In this section we find the classical constraints between gauge singlet operators in the $A+4Q + N\barQ$ model, along with the coefficients in the dynamically generated superpotential. It is useful to consider a particular non-trivial solution of the $D$ flatness conditions.

\subsection{$D$-Flat Directions} \label{sec:dflatness}

The auxiliary gluon scalar fields have interactions from the K\"ahler potential given by $V=\frac{1}{2} D^a D^a$, where
\begin{equation}
D^a = - g \left( Q^{\star\alpha}_i (T^a_{\yF})^\beta_\alpha Q^i_\beta + \barQ^{\star j}_\alpha (T^a_{\yFb})_\beta^\alpha \barQ_j^\alpha + A^{\star \beta\alpha} (T^a_{\yA})_{\alpha\beta}^{\delta\epsilon} A_{\delta\epsilon} \right). \label{eq:Dflatness}
\end{equation}
Ground state solutions are given by $D^a D^a = 0$.
\Eqref{eq:Dflatness} can be simplified by replacing $T_{\ysub{\yFb}}$ and $T_{\ysub{\yA}}$ with $T^a_{\ysub{\yF}}$:
\begin{align}
(T^a_{\ysub{\yFb}})^\alpha_\beta = -(T^a_{\ysub{\yF}})^\beta_\alpha &,&
(T^a_{\ysub{\yA}})_{\alpha\beta}^{\delta\epsilon} = (T^a_{\ysub{\yF}})^\delta_\alpha \delta^\epsilon_\beta + \delta^\delta_\alpha (T^a_{\ysub{\yF}})^\epsilon_\beta  .
\end{align}
With this substitution, we may write $D^a$ as
\begin{eqnarray}
%D^a &=& - g \left( Q^{\star\alpha}_i Q^i_\beta (T^a_{\yF})^\beta_\alpha  - \barQ_j^\alpha \barQ^{\star j}_\beta (T^a_{\yF})^\beta_\alpha  + 2 A^{\star \alpha\gamma} A_{\gamma\beta } (T^a_{\yF})^\beta_\alpha \right) \\
D^a &=& -g \left( Q^{\star\alpha}_i Q^i_\beta  - \barQ_j^\alpha \barQ^{\star j}_\beta   + 2 A^{\star \alpha\gamma} A_{\gamma\beta } \right) (T^a_{\ysub{\yF}})^\beta_\alpha .
\end{eqnarray}
The indices $i$ and $j$ refer to $SU(4)_L$ and $SU(N)_R$, respectively, while $\alpha$, $\beta$ and $\gamma$ correspond to the gauge group.
The generators $T^a_{\ysub{\yF}}$ span the set of traceless $N\times N$ matrices, so if the fields satisfy
\begin{equation}
Q^{\star\alpha}_i Q^i_\beta  - \barQ_j^\alpha \barQ^{\star j}_\beta + 2 A^{\star \alpha\gamma} A_{\gamma\beta } = \rho \delta^\alpha_\beta  \label{eq:dappx}
\end{equation}
for any constant $\rho$, then $D^a=0$. 
It is useful to define the matrices $d$, $\bar d$, and $d_A$ as follows:
%\begin{eqnarray}
%d^\alpha_\beta &=& Q^{\star\alpha}_i Q^i_\beta \\
%\bar d^\alpha_\beta &=& \barQ_j^\alpha \barQ^{\star j}_\beta \\
%(d_A)^\alpha_\beta &=& A^{\star \alpha\gamma} A_{\gamma\beta },
%\end{eqnarray}
\begin{align}
d^\alpha_\beta = Q^{\star\alpha}_i Q^i_\beta &,&
\bar d^\alpha_\beta = \barQ_j^\alpha \barQ^{\star j}_\beta &,&
(d_A)^\alpha_\beta = A^{\star \alpha\gamma} A_{\gamma\beta },
\end{align}
so that \eqref{eq:dappx} can be written as
\begin{equation}
d^\alpha_\beta - \bar d^\alpha_\beta + 2(d_A)^\alpha_\beta = \rho \delta^\alpha_\beta .
\end{equation}
Each $d$ term defined above is invariant under the $SU(4)_L\times SU(N)_R$ flavor transformations.

By rotating the $SU(N)$ color basis, it is possible to block-diagonalize the matrix $A$ such that the only non-zero entries are $A_{12} = - A_{21} = \sigma_1$, $A_{34}= -A_{43} = \sigma_2$, etc. For even $SU(N=2m)$, this continues until $\sigma_m = A_{N-1, N}$. In this basis, the $d_A$ matrix is diagonal and equal to 
\begin{equation}
(d_A)^\alpha_\beta = \text{Diag}(|\sigma_1|^2, |\sigma_1|^2, |\sigma_2|^2, |\sigma_2|^2, \ldots, |\sigma_m|^2, |\sigma_m|^2 ),
\end{equation}
with $\Pf A = \sigma_1 \sigma_2 \ldots \sigma_m$.
For odd $N = 2m +1$, $\sigma_m = A_{N-2, N-1}$, and $A_{j N} = 0$ for all $j = 1 \ldots N$. The $d_A$ matrix is again diagonal, but with $(d_A)^N_N = 0$.
\begin{equation}
(d_A)^\alpha_\beta = \text{Diag}(|\sigma_1|^2, |\sigma_1|^2, |\sigma_2|^2, |\sigma_2|^2, \ldots, |\sigma_m|^2, |\sigma_m|^2, 0 ).
\end{equation}
The Pfaffian, $\Pf A$, is not defined for odd-dimensional matrices.

It is not generally possible to simultaneously diagonalize $d_A$, $d$, and $\bar d$.
%Alternatively, we could choose to diagonalize $\bar d$ or $d$ instead: however, we do not generally expect that $d$, $\bar d$ and $d_A$ can be simultaneously diagonalized. 
This is a departure from \susy\ QCD: in this case, if $\bar d$ is diagonal, then $d^\alpha_\beta = \bar d^\alpha_\beta + \rho\delta^\alpha_\beta$ must also be diagonal.
Once $d_A$ is added, this condition is relaxed.

\subsection{Special Cases}

In this section we consider the $\ev{\phi}\gg \Lambda$ limit along particular flat directions in which $d_A$, $d$, and $\bar d$ happen to be diagonal. Let us begin with the $N=2m$ case:
\begin{align}
A = \left( \begin{array}{c c c c c c c } 0 &\sigma_1 & &&&& \\ -\sigma_1 & 0 & &&&& \\ && 0 & \sigma_2 & && \\ && -\sigma_2 & 0 & && \\ &&&& \ddots&& \\ &&&&& 0 & \sigma_m  \\ &&&&& -\sigma_m & 0 \end{array} \right), &&
Q = \left( \begin{array}{c c c c} v_1 &  0 & & \\0 & v_2 & 0 & \\& 0 & v_3 & 0 \\& & 0 & v_4 \\ 0 & 0 & 0 & 0 \\\vdots & & & \vdots \\0 & 0 & 0 & 0 %\\q_1 & q_2 & q_3 & q_4 
\end{array} \right),&&
\barQ = \left( \begin{array}{c c c c c} \bar{v}_1 & 0  & & & \\0 & \bar{v}_2 &  & & \\& 0 & \ddots &  & \\& &  & \ddots & 0 \\  &  &  & 0 & \bar{v}_N  \end{array} \right) .%&~~
\label{eq:dflatnessA}
\end{align}
In this vacuum, the matrices $d_A$, $d$ and $\bar d$ are:
\begin{eqnarray}
d_A &=& \text{Diag}\left(\abs{\sigma_1}^2, \abs{\sigma_1}^2, \abs{\sigma_2}^2, \abs{\sigma_2}^2, \ldots , \abs{\sigma_m}^2, \abs{\sigma_m}^2 \right) \\
d &=& \text{Diag}\left(\abs{v_1}^2, \abs{v_2}^2, \abs{v_3}^2, \abs{v_4}^2, 0,  \ldots , 0 \right) \\
\bar d &=& \text{Diag}\left(\abs{\bar v_1}^2, \abs{\bar v_2}^2, \abs{\bar v_3}^2, \ldots , \abs{\bar v_{N-1}}^2, \abs{\bar v_N}^2 \right),
\end{eqnarray}
subject to the constraint
%\begin{eqnarray}
%\rho &=& |v_1|^2 - |\bar v_1|^2 + 2|\sigma_1|^2 \\
%&\vdots& \nonumber\\
%\rho &=& |v_4|^2 - |\bar v_4|^2 + 2|\sigma_2|^2 \\
%\rho &=& - |\bar v_5|^2 + 2|\sigma_3|^2 \\
%&\vdots& \nonumber\\
%\rho &=&- |\bar v_{N-1}|^2 + 2|\sigma_m|^2 \\
%\rho &=&- |\bar v_N|^2 + 2|\sigma_m|^2 .
%\end{eqnarray}
\begin{equation}
d^\alpha_\alpha - \bar d^\alpha_\alpha + 2(d_A)^\alpha_\alpha = \rho .
\end{equation}
In the classical limit, the gauge-invariant operators are
\begin{align}
J= \left( \begin{array}{c c c c} \bar v_1 v_1 &  0 & & \\0 & \bar v_2 v_2 & 0 & \\& 0 & \bar v_3 v_3 & 0 \\& & 0 & \bar v_4 v_4 \\ 0 & 0 & 0 & 0 \\\vdots & & & \vdots \\0 & 0 & 0 & 0 %\\ \bar v_N q_1 &\bar v_N q_2 & \bar v_N q_3 & \bar v_N q_4 
\end{array} \right)
&,~~~
%K= \left( \begin{array}{c c c c c c c c} 0 &\hat\sigma_1 & &&&&&0 \\ -\hat\sigma_1 & 0 & &&&&& \\ && 0 & \hat\sigma_2 & &&& \\ && -\hat\sigma_2 & 0 & &&& \\ &&&& \ddots&&& \\ &&&&& 0 & \hat\sigma_M & 0 \\ &&&&& -\hat\sigma_M & 0 & 0 \\ 0&&&&& 0 & 0 & 0 \end{array} \right).
K= \left( \begin{array}{c c c c c c c} 0 &\hat\sigma_1 & &&&& \\ -\hat\sigma_1 & 0 & &&&& \\ && 0 & \hat\sigma_2 & && \\ && -\hat\sigma_2 & 0 & && \\ &&&& \ddots&& \\ &&&&& 0 & \hat\sigma_m  \\ &&&&& -\hat\sigma_m & 0  \end{array} \right), \\
%\end{align}
%%\begin{eqnarray}
%%X^i &=& \sigma_1 \sigma_2 \ldots \sigma_M q_i \\
%%%Y_1 &=& \sigma_1 v_3 v_4 \sigma_3 \ldots \sigma_M q_2 \\
%%%Y_2 &=& -\sigma_1 v_3 v_4 \sigma_3 \ldots \sigma_M q_1 \\
%%%Y_3 &=& v_1 v_2 \sigma_2 \sigma_3 \ldots \sigma_M q_4 \\
%%%Y_4 &=& -v_1 v_2 \sigma_2 \sigma_3 \ldots \sigma_M q_3 \\
%%%Y_i &=& \left\{\begin{array}{c c} \sigma_1 v_3 v_4 \sigma_3 \ldots \sigma_M q_2  & i=1 \\  -\sigma_1 v_3 v_4 \sigma_3 \ldots \sigma_M q_1 & i=2 \\ v_1 v_2 \sigma_2 \sigma_3 \ldots \sigma_M q_4 & i=3 \\ -v_1 v_2 \sigma_2 \sigma_3 \ldots \sigma_M q_3 & i=4 \end{array} \right. \\
%%Y_i &=& \left\{\begin{array}{c c} i=1:& \sigma_1 v_3 v_4 \sigma_3 \ldots \sigma_M q_2  \\ i=2:&  -\sigma_1 v_3 v_4 \sigma_3 \ldots \sigma_M q_1 \\ i=3:& v_1 v_2 \sigma_2 \sigma_3 \ldots \sigma_M q_4  \\ i=4:& -v_1 v_2 \sigma_2 \sigma_3 \ldots \sigma_M q_3 \end{array} \right. \\
%%Z &=& \bar v_1 \ldots \bar v_N .
%%\end{eqnarray}
%\begin{align}
V \;=\; \left(\begin{array}{c c c c}
0 & V_{12} & 0 & 0 \\ 
-V_{12} & 0 & 0 & 0 \\ 
0 & 0 & 0 & V_{34} \\
0 & 0 &- V_{34} & 0 
\end{array} \right)
&,~~~
\begin{array}{rcl} 
U &=& \sigma_1 \sigma_2 \ldots \sigma_m \\
\ww &=& v_1 v_2 v_3 v_4 \sigma_3 \ldots \sigma_m \\
Z &=& \bar v_1 \bar v_2 \bar v_3 \ldots \bar v_N ,
\end{array} \label{eq:vvuz}
\end{align}
where we define 
\begin{align}
V_{12} ~\equiv~ (v_1 v_2) \sigma_2 \sigma_3 \ldots \sigma_m &\, ,&
V_{34} ~\equiv~ \sigma_1 (v_3 v_4) \sigma_3 \ldots \sigma_m &\, ,&
\hat\sigma_i ~\equiv~ \sigma_i \bar v_{2i-1} \bar v_{2i}
\end{align}
for $i=1 \ldots m$.

In the $N=2m+1$ case we add a row and column to $A$, with $A_{\alpha, N} = A_{N, \beta} = 0$ for all $\alpha$ and $\beta$. The form of $\barQ$ is left unchanged, but we add a nontrivial $N^{th}$ row to $Q^i_N$ with entries $q^i\neq 0$. With these modifications, the matrices $d_A$, $d$ and $\bar d$ become
\begin{eqnarray}
d_A &=& \text{Diag}\left(\abs{\sigma_1}^2, \abs{\sigma_1}^2, \abs{\sigma_2}^2, \abs{\sigma_2}^2, \ldots , \abs{\sigma_m}^2, \abs{\sigma_m}^2, 0 \right) \\
d &=& \text{Diag}\left(\abs{v_1}^2, \abs{v_2}^2, \abs{v_3}^2, \abs{v_4}^2, 0,  \ldots , 0, \sum_i |q_i|^2 \right) \\
\bar d &=& \text{Diag}\left(\abs{\bar v_1}^2, \abs{\bar v_2}^2, \abs{\bar v_3}^2, \ldots , \abs{\bar v_{N-1}}^2, \abs{\bar v_N}^2 \right),
\end{eqnarray}
and the gauge-invariant operators are
\begin{align}
J= \left( \begin{array}{c c c c} \bar v_1 v_1 &  0 & & \\0 & \bar v_2 v_2 & 0 & \\& 0 & \bar v_3 v_3 & 0 \\& & 0 & \bar v_4 v_4 \\ 0 & 0 & 0 & 0 \\\vdots & & & \vdots \\0 & 0 & 0 & 0 \\ \bar v_N q_1 &\bar v_N q_2 & \bar v_N q_3 & \bar v_N q_4 
\end{array} \right)
&,~~~
K= \left( \begin{array}{c c c c c c c c} 0 &\hat\sigma_1 & &&&&&0 \\ -\hat\sigma_1 & 0 & &&&&& \\ && 0 & \hat\sigma_2 & &&& \\ && -\hat\sigma_2 & 0 & &&& \\ &&&& \ddots&&& \\ &&&&& 0 & \hat\sigma_M & 0 \\ &&&&& -\hat\sigma_M & 0 & 0 \\ 0&&&&& 0 & 0 & 0 \end{array} \right), \label{eq:jkodd}\\
\begin{array}{rcl}
X^i &=& \sigma_1 \sigma_2 \ldots \sigma_M q_i \\
Z &=& \bar v_1 \ldots \bar v_N 
\end{array} &,~~~
Y_i ~= \left\{\begin{array}{c c} i=1:& \sigma_1 v_3 v_4 \sigma_3 \ldots \sigma_M q_2  \\ i=2:&  -\sigma_1 v_3 v_4 \sigma_3 \ldots \sigma_M q_1 \\ i=3:& v_1 v_2 \sigma_2 \sigma_3 \ldots \sigma_M q_4  \\ i=4:& -v_1 v_2 \sigma_2 \sigma_3 \ldots \sigma_M q_3 \end{array} \right. . \label{eq:xyz}
\end{align}

\paragraph{Classical constraints}

The dynamically generated superpotential has the form $W \sim A^{N-2} Q^4 \bar Q^N$. For odd $N$, there are three ways to contract the gauge indices:
\begin{eqnarray}
W_d &=& \frac{\alpha}{\Lambda^b} \bigg( X^i Y_i Z + \beta_1 \epsilon_{i_1 \ldots i_4} \epsilon^{j_1 \ldots j_N} X^{i_1} (K_{j_1 j_2} \ldots K_{j_{N-4} j_{N-3} } ) J_{j_{N-2}}^{i_2} J_{j_{N-1}}^{i_3} J_{j_{N}}^{i_4} \nonumber\\&&\ + \beta_2 \epsilon^{j_1 \ldots j_N} Y_i  (K_{j_1 j_2} \ldots K_{j_{N-2} j_{N-1} } ) J_{j_N}^i  \bigg), \label{eq:superp:gen}
\end{eqnarray}
while for even $N$ there are five terms:
\begin{eqnarray}
W_d &=& \frac{\alpha}{\Lambda^b} \bigg( U \ww Z + \gamma_1 \epsilon_{i_1 \ldots i_4} V^{i_1 i_2} V^{i_3 i_4} Z  + \gamma_2 \epsilon_{j_1 \ldots j_N} \epsilon_{i_1 \ldots i_4} U \left(K_{j_1 j_2} \ldots K_{j_{N-5}, j_{N-4} } \right)\left(J_{j_{N-3}}^{i_1} \ldots J_{j_{N}}^{i_4}\right) \nonumber\\&&\ 
+ \gamma_3 \epsilon_{j_1 \ldots j_N} \epsilon_{i_1 \ldots i_4} V^{i_1 i_2} \left(K_{j_1 j_2} \ldots K_{j_{N-3}, j_{N-2} } \right)\left(J_{j_{N-1}}^{i_3}  J_{j_{N}}^{i_4}\right) 
+ \gamma_4 \ww \Pf K \bigg).
\end{eqnarray}
The relationships between the coefficients are determined by matching the equations of motion from $W_d$ to the classical constraints on the operators.

%For example, any permutation of a $K_{j_a j_b}$ with another $K_{j_c j_d}$ adds a factor of $(-1)^2=1$ from the permutations $(j_a \leftrightarrow j_c)$ and $(j_b \leftrightarrow j_d)$: an antisymmetric product $\epsilon \dot K^m$ will therefore include $m!$ duplicate terms from the permutations of the various $K$ with each other. Permutations such as $(j_a \leftrightarrow j_b)$ add a negative sign from the $\epsilon$ tensor: however, because $K$ is antisymmetric, $K_{j_b j_a} = -K_{j_a j_b}$, so that this is also a duplicate term. This adds a factor of $2^m$ to the product $\epsilon \dot K^m$, so that
%\begin{equation}
%\epsilon^{j_1 \ldots j_{2m} } K_{j_1 j_2} \ldots K_{j_{2m-1} j_{2m} } = 2^m m! K_{[j_1 j_2} \ldots K_{j_{2m-1} j_{2m}] }. 
%\end{equation}
%Antisymmetric products of $J$, such as $\epsilon_{i_1\ldots i_k} \epsilon^{j_1\ldots j_k} (J_{j_1}^{i_1} \ldots J_{j_k}^{i_k} )$, are invariant with respect to paired $i$ and $j$ permutations. Therefore, there are $k!$ equivalent terms for every contribution to $\epsilon \dot J^k \dot \epsilon$.
%
%Based on these considerations, we expect that
%\begin{eqnarray}
%W &=& \frac{\alpha}{\Lambda^b} \bigg[ X^i Y_i Z - \frac{\epsilon_{i_1 \ldots i_4} \epsilon^{j_1 \ldots j_N}}{ 2^{M-1} (M-1)! 3!  } X^{i_1} (K_{j_1 j_2} \ldots K_{j_{N-4} j_{N-3} } ) J_{j_{N-2}}^{i_2} J_{j_{N-1}}^{i_3} J_{j_{N}}^{i_4} \nonumber\\&&\ - \frac{\epsilon^{j_1 \ldots j_N} }{ 2^M M! } Y_i  (K_{j_1 j_2} \ldots K_{j_{N-2} j_{N-1} } ) J_{j_N}^i  \bigg]
%\end{eqnarray}
%We will consider the special case discussed above to verify this expectation.

In the classical limit for even $N$, it follows from \eqref{eq:vvuz} that
\begin{equation}
\Pf V = V_{12} V_{34} = (\sigma_1 \sigma_2 v_1 v_2 v_3 v_4)(\sigma_3 \ldots \sigma_m)^2 = U \cdot Z,
\end{equation}
for example, so that
\begin{equation}
\gamma_1 = - \frac{1}{2^2 2!}.
\end{equation}
Applying this technique to other products of gauge invariant operators, we find
\begin{align}
\gamma_2 = -\frac{1}{2^{m-2} (m-2)! 4!} &\, ,&
\gamma_3 = + \frac{1}{4 \cdot 2^{m-1} (m-1)!} &\, ,&
\gamma_4 = - 1.
\end{align}
For odd $N$ the relevant classical constraints have the form
\begin{eqnarray}
X^i Z &=& - \beta_2 \epsilon^{j_1 \ldots j_N} (K_{j_1 j_2} \ldots K_{j_{N-2} j_{N-1} } ) J^i_{j_N}  \label{eq:constraint:c1} \\ 
Y_i Z &=& -\beta_1 \epsilon_{i i_2 i_3 i_4} \epsilon^{j_1 \ldots j_N}  (K_{j_1 j_2} \ldots K_{j_{N-4} j_{N-3} } ) J_{j_{N-2}}^{i_2} J_{j_{N-1}}^{i_3} J_{j_{N}}^{i_4} . \label{eq:constraint:c2}
\end{eqnarray}
Based on Eqs.~(\ref{eq:jkodd}) and~(\ref{eq:xyz}), 
\begin{eqnarray}
% c_1 \epsilon^{j_1 \ldots j_N} (K_{j_1 j_2} \ldots K_{j_{N-2} j_{N-1} } ) J^i_{j_N} 
% &=&  c_1 \epsilon^{j_1 \ldots j_{N-1} N} (K_{j_1 j_2} \ldots K_{j_{N-2} j_{N-1} } ) J^i_{N}  \\
% &=& 2^{M} M! c_1 \hat\sigma_1 \hat\sigma_2 \ldots \hat\sigma_M (\bar v_N q_i) \\
X^i Z &=& (\sigma_1 \ldots \sigma_m q_i)(\bar v_1 \ldots \bar v_N) \\
Y_i Z &=& (\sigma_1 v_3 v_4 \sigma_3 \ldots \sigma_m q_2)(\bar v_1 \ldots \bar v_N)  ,
\end{eqnarray}
which when matched to the corresponding products of $J$ and $K$ imply that
\begin{align}
\beta_1 = -\frac{1}{2^{m-1} (m-1)! 3!} &\, ,&
\beta_2 = -\frac{1}{2^m m!} .
\end{align}

In both cases the overall constant $\alpha$ has no effect on the equations of motion, and cannot be calculated from the classical constraints.

\bibliography{prodsconf}

\end{document}